\documentclass[12pt,a4paper]{article}
\usepackage{jheppub}
\usepackage{hyperref}

\usepackage{mathtext}         

\usepackage{amsmath,amssymb} 
\usepackage{amsfonts}
\usepackage{epsfig}

\def\al{\alpha}
\def\ap{\alpha'}
\def\bp{\beta'}
\def\bt{\beta}

\def\ga{\gamma}

\def\de{\delta}
\def\({\left(}
\def\){\right)}

\newcommand{\Ga}[1]{\Gamma\left({#1}\right)}
\newcommand{\scr}{\scriptscriptstyle}

\newcommand{\be}{\begin{equation}}
\newcommand{\ee}{\end{equation}}
\newcommand{\bea}{\begin{eqnarray}}
\newcommand{\eea}{\end{eqnarray}}
\newcommand{\ba}{\begin{eqnarray*}}
\newcommand{\ea}{\end{eqnarray*}}

\newcommand{\Fh}[2]{\,{}_#1F_#2}
\newcommand{\Fs}[3]{\!\!\left[\begin{array}{c}#1\,;\\#2\,;\end{array}#3\right]}

\newcommand{\Fy}[2]{\Fs{#1}{#2}{y}}

\newcommand{\FfP}[2]{\Fs{#1}{#2}{\frac{p^2}{4m^2}}}

\newcommand{\Fz}[2]{\Fs{#1}{#2}{z}}
\newcommand{\FzZ}[2]{\Fs{#1}{#2}{z_1}}
\newcommand{\FeZ}[2]{\Fs{#1}{#2}{z_1+z_2}}
\newcommand{\FEZ}[2]{\Fs{#1}{#2}{z_1}}
\newcommand{\Fzm}[2]{\Fs{#1}{#2}{1-z}}

\newcommand{\Frr}[2]{\Fs{#1}{#2}{\frac{r_{23}}{r_{123}} }}
\newcommand{\Frd}[2]{\Fs{#1}{#2}{\frac{r_{2}}{r_{123}} }}
\newcommand{\Frt}[2]{\Fs{#1}{#2}{\frac{r_{3}}{r_{123}} }}

\newcommand{\FRD}[2]{\Fs{#1}{#2}{\frac{x(1-y)}{y(1-x)} }}
\newcommand{\FRT}[2]{\Fs{#1}{#2}{\frac{x}{y} }}
\newcommand{\FRR}[2]{\Fs{#1}{#2}{\frac{ r_3-r_{23}}{r_{123}-r_{23}} }}
\newcommand{\FOO}[2]{\Fs{#1}{#2}{\frac{r_{234}}{r_{1234}} }}

\newcommand{\Fx}[2]{\Fs{#1}{#2}{x}}

\newcommand{\Fmr}[2]{\Fs{#1}{#2}{\frac{r_j}{r_{12}}}}

\newcommand{\Fwz}[2]{\Fs{#1}{#2}{\frac{w-z}{1-z}}}

\newcommand{\Fup}[2]{\Fs{#1}{#2}{\frac{p^2}{m^2}}}

\newcommand{\For}[2]{\Fs{#1}{#2}{1-\frac{r_j}{r_{12}}}}

\renewcommand{\thesection}{\normalsize \arabic {section}}
\begin{document}

\begin{titlepage}

\begin{flushright}
\end{flushright}

\vspace*{0.2cm}
\begin{center}
{\Large {Functional reduction of one-loop Feynman integrals with arbitrary masses
}}\\[2 cm]
\end{center}

\begin{center}
{\bf  O.V.~Tarasov}
\vspace{0.5cm}
\\   

\it Joint Institute for Nuclear Research,\\
      141980 Dubna, Russian Federation \\
     E-mail: {\tt otarasov@jinr.ru}
\end{center}

\vspace*{1.0cm}

\begin{abstract}

A method of  functional reduction for the dimensionally regularized 
one-loop Feynman 
integrals with massive propagators is described in detail.

The method is based on a repeated application of the functional
relations proposed by the author. Explicit formulae are given  for  
reducing one-loop scalar integrals  to  a simpler ones, the arguments 
of which are the ratios of polynomials in the  masses and 
kinematic invariants.
We show that a general scalar $n$-point integral, 
depending on $n(n+1)/2$ generic masses and kinematic variables,
can be expressed as a linear combination of integrals
depending only on $n$ variables.
The latter integrals are given explicitly in terms of  
hypergeometric functions of $(n-1)$ dimensionless 
variables.
Analytic expressions for the 2-, 3- and 4-point integrals, that
depend on the minimal number of variables, were also obtained 
by solving the dimensional recurrence relations. The resulting
expressions for these integrals are given in terms of 
  Gauss'  hypergeometric function $_2F_1$,  the Appell function $F_1$
 and the hypergeometric Lauricella - Saran  function $F_S$.
A modification of the functional reduction  procedure for some special
values of kinematical variables is considered.

\end{abstract}

\end{titlepage}
\tableofcontents

\section{Introduction}

Feynman integrals play an important role in making precise
perturbative predictions in quantum field theory and statistical physics.
Theoretical predictions for  experiments at the LHC
\cite{Aad:2012tfa,Chatrchyan:2012xdj} as well as
at future colliders such as the FCC \cite{FCC:2018byv}
demand  knowledge of  precise
radiative corrections. Precise experimental measurements
are to be interpreted with  sufficient precision of theoretical 
predictions \cite{Heinrich:2020ybq}.
The complexity of the evaluation of such radiative corrections is related, in
particular, to the difficulties in calculating integrals corresponding to
Feynman diagrams with many external legs depending on many kinematical
variables.
Purely numerical evaluation of such integrals sometimes cannot provide 
sufficiently high precision within reasonable computer time.
Problems of numerical evaluation of the one-loop integrals
have been considered, for example, in refs. \cite{Passarino:1978jh},
\cite{vanOldenborgh:1989wn},
\cite{Ferroglia:2002mz}, \cite{Ellis:2011cr}.
Numerical instability in evaluating one-loop
scalar integrals near exceptional momentum configurations
was addressed in refs. \cite{Campbell:1996zw}, 
\cite{Jegerlehner:2002es}, \cite{Giele:2004ub}.

At present, there are many various methods of evaluating Feynman integrals.
Some methods are based on the fact that these integrals 
 are functions of continuous variables - scalar products of external 
momenta  and masses, and are also functions of discrete parameters - 
powers of propagators as well as space - time dimension parameter $d$.
External kinematic invariants and squared masses were used to derive
differential equations \cite{Kotikov:1990kg}
(see also reviews \cite{Golubeva:1974}, \cite{Argeri:2007up}).
Space-time dimension $d$ and powers of propagators were used
to derive difference equations  \cite{Kazakov:1983pk}, \cite{Tarasov:1996br},
\cite{Laporta:2000dsw}
for these integrals. 
Then the results  for integrals are obtained by solving  these equations.
Practical application of the method of differential equation  and 
methods based on recurrence relations to evaluating high-order, multi-leg
Feynman diagrams, clearly demonstrate
the need for further improvements and development of
methods for solving  differential and recursion relations.

On the other hand, it is possible to extend the applicability of these 
methods by combining them with other approaches.  For instance,
these methods can be used in combination with approach
proposed in refs.~\cite{Tarasov:2008hw}, \cite{Tarasov:2015wcd},
\cite{Tarasov:2019mqy}.
In ref. \cite{Tarasov:2008hw}, a new type of relations between 
Feynman integrals, namely functional
relations has been discovered.
In ref. \cite{Tarasov:2015wcd} a simple method has been 
proposed for deriving
functional relations applicable to integrals corresponding 
to Feynman diagrams with any number of loops and legs.
Using these relations a method of functional
reduction was formulated   and
applied to several massless integrals in ref. \cite{Tarasov:2019mqy}.
This method allows one to express the integral of interest in terms
of integrals with fewer variables.
In general, the latter integrals will be easier to evaluate
by the above mentioned methods than the original integral.
 
Integrals  appearing in the final results of  functional 
reduction have two important features.
Firstly, that they depend on the minimal number of variables 
(MNV) and, secondly, these variables are the ratios of 
Gram determinants.

As for our representation of integrals in terms of functions that depend
explicitly on ratios of the Gram determinants, we would like to mention 
refs.~\cite{vanOldenborgh:1989wn},\cite{Giele:2004ub} where 
the importance of representing kinematic dependence of integrals
in terms of ratios of the Gram determinants  was  demonstrated.
As the authors have shown, such a representation
turns out to be  useful for the stability of numerical calculation of
integrals.

 The primary purpose of the paper is to apply the method of functional
 reduction to the scalar one - loop integrals that depend
 on arbitrary kinematic variables and masses. 

The article is organized as follows.
In Section 2, we will briefly describe the method for deriving
functional relations given in ref.~\cite{Tarasov:2015wcd}.
In Section 3,  we describe the  method of functional
reduction proposed in  ref.~\cite{Tarasov:2019mqy}.
In Section 4, the functional reduction of the 2-point integral
is considered.
In Section 5, a two-step functional reduction of the 
integral corresponding to a 3-point Feynman diagram is
described. The Feynman parameter representation and dimensional 
recurrence relations for the integrals arising
at the final stage of the functional reduction are given.
We present an analytic result derived by
a use of the dimensional recurrence relation as well as
a result in terms of a double hypergeometric series,
obtained by expanding the Feynman parameter integral.

In Section 6, we propose three-step functional
reduction procedure for the 4-point integral.
Solving the dimensional recurrence relation, we obtained an 
analytic result for the integral depending on the MNV.
Also, representation of this integral in terms of
a triple hypergeometric series is given.

In Section 7, we describe derivation of functional
relations of a four-step reduction procedure for the 5-point integral.
We also give here the Feynman parameter representation
of the 5-point integral depending on the MNV as well as
a dimensional recurrence relation for this integral.
Using  a parametric representation of the
integral, we expressed it as a fourfold hypergeometric series.

In Section 8, we describe 5 steps of  the functional reduction
of a 6-point integral. The Feynman parameter representation
and a dimensional recurrence relation for the integral
with the MNV are given.  Using a parametric representation,
we expressed the integral as a multiple hypergeometric series.

In Section 9, we describe a modification  of the functional
reduction method for  integrals depending on special
values of kinematical variables and 
present analytic results for these integrals.

In Section 10, a general method is proposed for obtaining the 
final formula of the functional reduction for an arbitrary
one-loop $n$ - point integral. A parametric representation is also
given  for  the $n$-point integrals depending on the MNV. Using
 this parametric representation,  we obtain a  representation of the integral 
in terms of multiple hypergeometric series.

We offer some concluding remarks in Section 11.
Finally in the Appendix, we give  formulae for kinematic determinants and
hypergeometric functions that were used in our paper.

\section{Algebraic relation between propagators}

We consider one-loop scalar integral in general dimension  $d$ 
corresponding to a Feynman diagram with $n$ external lines
and $n$ internal propagators  with  
arbitrary masses $m_i$ and external momenta 
\begin{eqnarray}
&&
I_n^{(d)}( \{m_j^2 \}; \{s_{ik}\})  =
\frac{1}{i\pi^{d/2}}\int
\frac{d^dk_1}{D_1 {\ldots} D_n}
\end{eqnarray}
where the massive propagators have the form
\begin{equation}
D_j=(k_1-p_j)^2 - m_j^2+i\eta.
\end{equation} 
In what follows we will omit the $i\eta$ term assuming that 
all masses have such a correction.
The propagators and momenta are labeled as in Figure 1.
\vspace{0.5cm}
\begin{figure}[h]
\begin{center}
\includegraphics[scale=0.9]{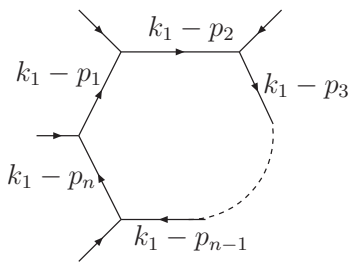}
\end{center}
\caption{The generic $n$-point one-loop graph.}
\end{figure}
As was shown in ref.~\cite{Tarasov:2015wcd} the functional
relations for these integrals can be derived
from  the following algebraic relation
between the products of propagators
\begin{equation}
\prod_{r=1}^{n}\frac{1}{D_r} = \frac{1}{D_{0}}
\sum_{r=1}^n x_r {\prod_{\substack{j=1
\\ 
j \neq r} }^n} \frac{1}{D_j}.
\label{usual_props}
\end{equation}
We assume that $k_1$ is an arbitrary momentum and $p_j$  
correspond to external momenta.
The proceeding equation is satisfied if $p_0$, $m_0^2$
and  $x_j$, ($j=1, ...  n$) 
are chosen to satisfy the system of equations.
In order to obtain such a system we proceed as follows. We
multiply both sides of the Eq.~(\ref{usual_props}) by the product
$\prod_{j=0}^{n} D_j$ and get 
\begin{equation}
D_{0}= \sum_{r=1}^{n} x_r D_r,
\end{equation}
or
\begin{equation}
k_1^2-2k_1p_{0}+p_{0}^2-m_0^2=\sum_{r=1}^n x_r(k_1^2-2k_1p_r+p_r^2 -m_r^2).
\label{ini_equ}
\end{equation}
It is assumed that $k_1$ will be an  integration momentum and
$p_j$, $x_r$ do not depend on it. Differentiating both sides of 
Eq.(\ref{ini_equ})
with respect to $k_{1\mu}$, one gets a linear equation in $k_1$ from which
two equations follow
\begin{equation}
1= \sum_{r=1}^nx_r,
\label{sumX}
\end{equation}
\begin{equation}
p_{0}= \sum_{j=1}^n x_{j} p_j.
\label{pnp1}
\end{equation}
Substituting Eqs.~(\ref{sumX}),(\ref{pnp1}) into Eq.~(\ref{ini_equ})
yields the following  equation
\begin{equation}
m_{0}^2-\sum_{k=1}^n x_{k} m_k^2 +\sum_{j=2}^{n}\sum_{l=1}^{j-1}
x_{j}x_{l}s_{lj}=0,
\label{kwadraticY}
\end{equation}
where kinematic invariants $s_{ij}$ are defined as
\begin{equation}
s_{ij}=s_{i,j}=(p_i-p_j)^2.
\label{sij}
\end{equation}
Solving Eq.~(\ref{sumX}) for one of the parameters $x_{j}$ and then substituting 
this  solution into Eq.~(\ref{kwadraticY}) gives a quadratic equation
for the remaining parameters $x_{i}$. This quadratic equation can be solved
with respect to  one of the parameters $x_{j}$. Thus, the solution of  the system
of equations  (\ref{sumX}), (\ref{kwadraticY}) depends on
$(n-2)$ of the remaining arbitrary parameters $x_{i}$ and one arbitrary mass  $m_0$.

Integrating   algebraic relation (\ref{usual_props})
over momentum { $k_1$}
yields functional equation for a general one-loop $n$-point integral
\begin{eqnarray}
&&I_n^{(d)}(\{ m_r^2\};\{s_{ik}\} ) =\sum_{j=1}^n x_j
\left.I_n^{(d)}(\{ m_r^2\};\{s_{ik}\} )\right|_{m_j^2\rightarrow m_0^2,
s_{jk}\rightarrow s_{0k}}.
\label{funcura_arb_n}
\end{eqnarray}
This equation will be our starting equation for deriving  relations for
the functional reduction
of integrals $I_n^{(d)}(\{ m_r^2\};\{s_{ik}\} )$.
In the next sections, we will consider in detail the derivation
of functional relations for 
reducing integrals $I_2^{(d)}$,...,$I_6^{(d)}$.

\section{Method of functional reduction}

By choosing arbitrary parameters $x_j$, $m_0^2$,  we can try to
express the integral of interest in terms of integrals with fewer
variables. If we manage to find such parameters we will 
actually solve the functional equation for the integral. 

A systematic method for  solving functional equations
for Feynman integrals  was presented in ref.~\cite{Tarasov:2019mqy}.
In a sense, this is a generalization of the method that is used 
to solve the usual Sincov's functional equation 
\cite{Sincov:1903a}, \cite{Sincov:1903b},
\cite{castillo2004functional}
\begin{eqnarray}
f(x,y)=f(x,z)-f(y,z).
\label{Sincov_equ}
\end{eqnarray}
By setting $z=0$ in this equation, we get a general solution 
\begin{equation}
f(x,y)= g(x)-g(y),
\end{equation}
where 
\begin{equation}
g(x)=f(x,0).
\end{equation}
I.e.  the function $f(x,y)$
is a combination  of its 'boundary values', which may be 
completely arbitrary.

As for solving functional equations for Feynman integrals,
the situation is much
more complicated here - too many variables are involved, too many functions .
For this reason, we used a computer to systematically  search for  
possible relationships between the  arguments of integrals
leading to a decrease in the number of variables of these integrals. 
To reduce the number of variables, we
impose the following simple conditions on the new variables
$s_{j0}$, $m_0^2$
\begin{eqnarray}
&&
   s_{0j}=0,~~s_{0j}-s_{0i}=0,~~s_{0j}\pm s_{ik}=0,
 	 ~~s_{0j} \pm m_0^2=0,~~m_j^2\pm m_0^2=0,
\nonumber \\
&&m_0^2=0,~~~~ s_{0j} \pm m_0^2 \pm m_k^2=0,~~~~~(i,j,k =1{\ldots} n).~~~~~~~~~~~~~~~
\label{constrains}
\end{eqnarray}
From the set of equations obtained by combining Eqs.~(\ref{constrains}),
(\ref{sumX}) and (\ref{kwadraticY}), we have formed various systems of equations
with $2,3,4,  $ etc.  equations in each system.
Solutions of these systems of equations and 
analysis of these solutions were performed using computer algebra 
system MAPLE. The number of these systems depends on $n$ and varied from 
$10^3$ to $10^6$. CPU execution time ranged from
a few  minutes to several hours. 
Many solutions of these equations have been found. Some of them lead to 
a simultaneous decrease in the number of variables
in all integrals on the right-hand side of the functional equation
(\ref{funcura_arb_n}).
In the following sections we will describe in detail how this method works.

\section{Functional reduction of the 2-point integral  $I_2^{(d)}$ }
We will start by considering a simple one-loop integral depending 
on arbitrary masses and external momentum
\begin{equation}
I_2^{(d)}(m_1^2,m_2^2;~s_{12})=
\int \frac{d^d k_1}{i \pi^{{d}/{2}}}
\frac{1}{[(k_1-p_1)^2-m_1^2]
         [(k_1-p_2)^2-m_2^2]}.
\end{equation}
Setting $n=2$ in the Eq.~(\ref{usual_props}), leads to an 
algebraic relation between the products of two propagators
 \begin{eqnarray}
\frac{1}{D_1D_2}=\frac{x_1}{D_0D_2}+\frac{x_2}{D_1D_0}.
\label{alg_rel_I2}
\end{eqnarray}
At $n=2$, according to (\ref{sumX})--(\ref{kwadraticY}),  parameters
$x_j$, $m_0^2$ and momentum $p_0$ in this equation must 
obey the following conditions
\begin{eqnarray}
&& x_1+x_2=1,~~~~~~p_0=x_1p_1 + x_2 p_2,
\nonumber \\
&& m_0^2-x_1m_1^2-x_2m_2^2 +x_1x_2s_{12}=0.
\label{I2_conditions}
\end{eqnarray}
Integrating algebraic relation (\ref{alg_rel_I2})
over momentum {$k_1$} yields
\begin{eqnarray}
 I_2^{(d)}(m_1^2,m_2^2; s_{12})=
 { x_1} 
 I_2^{(d)}(m_2^2,m_0^2; s_{20})
+{x_2} I_2^{(d)}(m_1^2,m_0^2; {s_{10}}).
\label{I2funcequ}
\end{eqnarray} 
By solving system of equations (\ref{I2_conditions})
for $x_1$, $x_2$, we get 
\begin{eqnarray}
&&
x_1 = \frac{m_2^2-m_1^2+s_{12}}{2s_{12}}
 \pm \frac{\sqrt{4s_{12}
 (m_0^2-r_{12})  }}{2 s_{12}},
~~~~~~~{x_2}=1-x_1.
\label{x1x2}
\end{eqnarray}
Both kinematic invariants $s_{10}$, $s_{20}$ can be expressed in terms of $x_1$ as
\begin{eqnarray}
&& s_{10}=(p_1-p_0)^2 =(1-x_1)^2s_{12}
=
\nonumber \\
&&~~~~~~~~~~~~~~~~~~~~~~~
m_1^2+{m_0^2} -2r_{12}\pm \frac{m_2^2-m_1^2-s_{12}}{2s_{12}} 
\sqrt{4s_{12}({m_0^2}-r_{12})},
\nonumber \\
&&
\label{S13_S23}
\nonumber
\\
&&{ 
s_{20}}=(p_2-p_0)^2 =x_1^2s_{12}=
\nonumber  \\
&&~~~~~~~~~~~~~~~~~~~~~~~~m_2^2 +{m_0^2}-2r_{12}
\pm \frac{m_2^2-m_1^2+s_{12}}{2s_{12}} 
\sqrt{4s_{12}({m_0^2}-r_{12})},
~~~
\end{eqnarray}
where
\begin{equation}
r_{12}=-\frac{\lambda_{12}}{g_{12}}=
\frac{2m_1^2m_2^2+2s_{12}m_1^2+2s_{12}m_2^2-m_1^4-m_2^4-s_{12}^2}
  {4s_{12}}.
\end{equation}  
The definitions of the determinants $\lambda_{12}$, $g_{12}$ are given
in Appendix.

Equation (\ref{I2funcequ}) strongly resembles Sincov's equation 
(\ref{Sincov_equ}).
By setting 
the only remaining arbitrary parameter $m_0^2$ to some special value,
one can try to reduce  the number of variables  simultaneously
for both integrals on the right side of Eq.~(\ref{I2funcequ}).
We will consider three different cases.

\begin{center}
{\it Case 1. $m_0^2=0$}\\
\end{center}
The most obvious choice is  to take $m_0^2=0$. 
Substituting this value into Eq.~(\ref{I2funcequ}),
we obtain
\begin{eqnarray}
 I_2^{(d)}(m_1^2,m_2^2; s_{12})=
 { \overline{x}_1} 
 I_2^{(d)}(m_2^2,0; { \overline{s}_{20}})
+{\overline{x}_2} I_2^{(d)}(m_1^2,0; {\overline{s}_{10}}),
\label{mm0_zero}
\end{eqnarray}
where
\begin{equation}
\overline{x}_{1,2}=\left. x_{1,2}\right|_{m_0^2=0},~~~~
\overline{s}_{01}=\left. s_{01} \right|_{m_0^2=0},~~~~
\overline{s}_{02}=\left. s_{02} \right|_{m_0^2=0}.
\end{equation}
The analytic expression for the integral  $I_2^{(d)}(m^2, 0;~p^2)$
is well known
(see refs. \cite{Bollini:1972bi}, \cite{Boos:1990rg})
\begin{equation}
I_2^{(d)}(m^2,0;~p^2) =
- \Gamma\left(1-\frac{d}{2}\right) m^{d-4} 
\Fh21\Fup{1,2-\frac{d}{2}}{\frac{d}{2}}.
\label{fermprop}
\end{equation}
Note that the $\varepsilon=(4-d)/2$ expansion
of the hypergeometric function  $_2F_1$ in Eq.~(\ref{fermprop}) is 
known to all orders in $\varepsilon$  \cite{Kalmykov:2006pu},
\cite{Kalmykov:2006hu}, \cite{Huber:2005yg}.  
Using Eq.~(\ref{mm0_zero}) one can easily obtain $\varepsilon$
expansion of the original integral $I_2^{(d)}(m_1^2,m_2^2; s_{12})$.
\begin{center}
{\it Case 2. $m_0^2=r_{12}$}\\
\end{center}
The second special value  of $m_0^2$, which leads to a 
simultaneous decrease
in the number of variables in both integrals 
on the right side of Eq.~(\ref{I2funcequ}), is
$m_0^2=r_{12}$.  In this case, the square roots in 
Eqs.~(\ref{x1x2}),(\ref{S13_S23}) vanish, and we get 
\begin{eqnarray}
 I_2^{(d)}(m_1^2,m_2^2; s_{12})=
 \kappa_{12} 
 I_2^{(d)}(r_{12},r_2; r_2-r_{12})
+{\kappa_{21}} I_2^{(d)}(r_{12},r_1; r_1-r_{12}),
\label{I2funcred}
\end{eqnarray}
where 
\begin{equation}
\kappa_{12}= \frac{\partial r_{12}}{\partial m_1^2},~~~~~
\kappa_{21}= \frac{\partial r_{12}}{\partial m_2^2},~~~~~~r_i=m_i^2.
\end{equation}
The analytic result for  integrals $I_2^{(d)}(r_{12},r_j,r_j-r_{12})$,
($j=1,2$) can be obtained either from a Feynman parameter representation
\begin{equation}
I_2^{(d)}(r_{12},r_j; r_j-r_{12})=
\Gamma\left(2-\frac{d}{2}\right)
\int_0^1\!\left[r_{12}+(r_j-r_{12})x_1^2\right]^{\frac{d}{2}-2}  dx_1 ,
\label{I2param}
\end{equation}
or by solving a dimensional recurrence relation
\begin{eqnarray}
&&(d-1)I_2^{(d+2)}(r_{12},r_j; r_j-r_{12})=
-2r_{12} I_2^{(d)}(r_{12},r_j; r_j-r_{12})
-I_1^{(d)}(r_j).~~~~
\label{I2dimrec}
\end{eqnarray}
In the latter case the result reads
\begin{eqnarray}
&&I_2^{(d)}(r_{12},r_j; r_j-r_{12})=
\frac{-\pi^{\frac32}  r_{12}^{\frac{d}{2}-2}}
{2\sin \frac{\pi d}{2} \Gamma\left(\frac{d-1}{2}\right)}
\sqrt{\frac{r_{12}}{r_{12}-r_j}} 
\nonumber\\
&& \nonumber \\
&&~~~~~~~~~~~~~~~~~~~~~+\frac{\pi}{2r_{12}} \frac{r_j^{\frac{d}{2}-1}}
{\sin \frac{\pi d}{2} \Gamma\left(\frac{d}{2}\right)}
\Fh21\Fmr{1,\frac{d-1}{2}}{\frac{d}{2}}.
\label{I2dimrec_solu}
\end{eqnarray}
It is valid for  $\left|r_{j}/r_{12}\right|<1$.
In order to solve the dimensional recurrence relation (\ref{I2dimrec}),
we used the method described in ref. \cite{Tarasov:2000sf}.

By changing variable in the integral
 (\ref{I2param}) and comparing the result with the integral
 representation of a $_2F_1$ function (\ref{F21intrep}),
 we find
\begin{eqnarray}
&&I_2^{(d)}(r_{12},r_j; r_j-r_{12})=
{r_{12}^{\frac{d}{2}-2}}{\Gamma\left(2-\frac{d}{2}\right)}
\Fh21\For{2-\frac{d}{2},\frac{1}{2}}{\frac32}.
\label{I2param_solu}
\end{eqnarray}
esult may also be obtained by expanding the integrand 
in Eq.~(\ref{I2param}) in powers of 
$z_j=1-r_j/r_{12}$,  assuming  $\left|z_j\right|<1$,
and then integrating with respect to  $x_1$ term by term.
Formula (\ref{I2dimrec_solu}) can be obtained from
Eq.~(\ref{I2param_solu}) by performing an analytic continuation 
of the hypergeometric function $_2F_1$.
\begin{center}
{\it Case 3. Combination of two equations}\\
\end{center}
The third reduction of integrals can be achieved in a
slightly more complicated way. First, we set $m_0^2=m_2^2$ 
in Eq.~(\ref{I2funcequ}) and  obtain
\begin{eqnarray}
I_2(m_1^2,m_2^2; s_{12})&=&
\frac{m_1^2-m_2^2}{s_{12}} 
I_2^{(d)}\left(m_1^2,m_2^2;~\frac{(m_1^2-m_2^2)^2}{s_{12}}\right)
\nonumber 
\\
&+& \frac{ s_{12}-m_1^2+m_2^2}{s_{12}}
I_2^{(d)}\left(m_2^2,m_2^2;~\frac{(s_{12}-m_1^2+m_2^2)^2}{s_{12}}
\right).
\label{mm0mm1}
\end{eqnarray}
Then we interchange masses $m_1^2 \leftrightarrow m_2^2$ in this formula  
and add the result to (\ref{mm0mm1}). 
Due to the invariance of $I_2^{(d)}$  
under this permutation, two terms in this sum 
having different signs  cancel out, so we get
\begin{eqnarray}
I_2(m_1^2,m_2^2;~s_{12})&=&
\frac{s_{12}+m_1^2-m_2^2}{2s_{12}} 
I_2^{(d)}\left(m_1^2,m_1^2;~\frac{(s_{12}+m_1^2-m_2^2)^2}{s_{12}}\right)
\nonumber 
\\
&+& \frac{ s_{12}-m_1^2+m_2^2}{2s_{12}}
I_2^{(d)}\left(m_2^2,m_2^2;~\frac{(s_{12}-m_1^2+m_2^2)^2}{s_{12}}
\right).
\label{I2fe_eqm}
\end{eqnarray}
The same result can be derived by setting $m_0^2=m_1^2$ in 
Eq.~(\ref{I2funcequ}) and then adding the obtained result to the 
result obtained by setting  $m_0^2=m_2^2$ in Eq.~(\ref{I2funcequ}).
An analytic expression for the  integral with equal masses is 
well known (see refs.~\cite{Bollini:1972bi}, \cite{Boos:1990rg}):
\begin{equation}
I_2^{(d)}(m^2,m^2; p^2)
 =m^{d-4}~
 \Gamma \left( 2- \frac{d}{2} \right) \Fh21\FfP{1,2-\frac{d}{2}}{\frac32}.
\label{I2mmmm}
 \end{equation}

Thus, we have presented three different possibilities of reducing 
the integral $I_2^{(d)}$ to a sum  of integrals with fewer
variables. Different reduction formulae can be used in
different kinematic domains.

Using Eqs.~(\ref{mm0_zero}), (\ref{I2funcred}), (\ref{I2fe_eqm}) 
one can easily find relations between integrals
that appeared in the right-hand sides of these equations.
For example, setting $m_1^2=r_{12}$, $m_2^2=r_j$,
$s_{12}=r_j-r_{12}$  in Eq.~(\ref{I2fe_eqm}), we get
\begin{equation}
I_2(r_{12},r_j; r_j-r_{12}) = I_2(r_j,r_j; 4(r_j-r_{12}))~~~~~(j=1,2).
\end{equation}

We conclude this section with a remark about the differences 
between integrals found in the three reduction procedures. 
In Eq.~(\ref{mm0_zero}), the arguments of  integrals on the 
right side depend on square 
roots of ratios of polynomials, while in Eqs.~(\ref{I2funcred}),
(\ref{I2fe_eqm}) arguments of integrals on the right side are 
just ratios of polynomials.  In all these cases, integrals were
expressed in terms of the hypergeometric function $_2F_1$. 
However, $\varepsilon =(4-d)/2$ expansion of  the functions
in Eq.~(\ref{mm0_zero}) technically slightly simpler than expansion
of the $_2F_1$ functions in Eqs.~(\ref{I2funcred}), (\ref{I2fe_eqm}).
The reason is that in the first case, parameters of the 
$_2F_1$ functions are integers  plus terms proportional to the
$\varepsilon$
while in the $_2F_1$ functions from Eqs.~(\ref{I2funcred}), (\ref{I2fe_eqm})
some parameters are half integers. The $\varepsilon$
expansion of the $_2F_1$ functions with half integer parameters contains
logarithms and polylogarithms depending on the square roots of the 
argument of the $_2F_1$  function \cite{Kalmykov:2006pu},
\cite{Kalmykov:2006hu}, \cite{Huber:2005yg},
while expansion of $_2F_1$ functions 
with integer parameters do not have such square roots.

Our preliminary study shows that a similar situation takes
place with integrals $I_3^{(d)}$ and $I_4^{(d)}$. Analytical 
results for these integrals involve the function $_2F_1$ as well 
as  more complicated hypergeometric functions. We expect 
that finding relationships between those
functions with integer and half integer parameters
will be helpful in performing the $\varepsilon$ expansion of integrals
$I_3^{(d)}$ and $I_4^{(d)}$.


\section{Functional reduction of the 3-point integral $I_3^{(d)}$}


Now we turn to a 3-point integral with arbitrary internal mass scales and 
arbitrary external momenta
\begin{equation}
I_3^{(d)}(m_1^2,m_2^2,m_3^2;s_{23},s_{13},s_{12})
=\frac{1}{i \pi^{d/2}} \int \frac{d^dk_1}{D_1D_2D_3}.
\label{i3_definition}
\end{equation}
Setting $n=3$ in equation (\ref{usual_props}) leads to an
algebraic relation for the  products of three propagators \cite{Tarasov:2015wcd}:
\begin{equation}
\frac{1}{D_1 D_2 D_3}= \frac{x_{\scr 1}}{D_0 D_2 D_3 }
+\frac{x_{2}}{D_1 D_0 D_3}+\frac{x_{ 3}}{D_1 D_2  D_0}.
\label{3prop_relation}
\end{equation}
Equation (\ref{3prop_relation}) holds if
\begin{equation}
p_0 = x_1p_1+x_2p_2+x_3p_3,
\label{I3_p0}
\end{equation}
and the parameters $m_0^2$, $x_j$ obey the following system of equations:
\begin{eqnarray}
&&x_{1}+x_{2}+x_{ 3}=1,
\nonumber \\
&&x_{ 1}x_{ 2}s_{12}+x_{1}x_{\scr 3}s_{ 13}+x_{2}x_{ 3}s_{ 23}
-x_{1}m_1^2-x_{ 2}m_2^2-x_{3}m_3^2
 +m_{ 0}^2=0.
\label{I3_x_m}
\end{eqnarray}
Integrating  (\ref{3prop_relation}) over
 momentum $k_1$ gives a functional relation for the one-loop
integral $I_3^{(d)}$ with arbitrary masses and kinematic 
variables:
\begin{eqnarray}
&&I_3^{(d)}(m_1^2,m_2^2,m_3^2;  s_{ 23},s_{ 13},s_{ 12})
\nonumber \\
&&~~~~~~~~ 
=x_{1}
I_3^{(d)}(m^2_{0},m_2^2,m_3^2;  s_{ 23},s_{ 03},s_{02})
\nonumber \\
&&~~~~~~~~ +x_2 
I_3^{(d)}(m_1^2,m_{0}^2,m_3^2;  {s}_{03},
{s}_{13}, s_{ 01})
\nonumber \\
&&~~~~~~~~
+x_{3} I_3^{(d)}(m_1^2,m_2^2,m_{0}^2;  s_{02},{s}_{01},{s}_{12}),
\label{feI3massive}
\end{eqnarray}
Now our aim is to find the values of $m_0^2$, $x_j$ ($j=1,2,3$),
leading to a simultaneous reduction in the number 
of variables in all integrals on the right side of Eq.~(\ref{feI3massive}). 
Equation (\ref{feI3massive}) will be our starting point at all steps
of  the functional reduction.

\subsection{Functional reduction procedure}
Functional reduction of the 3-point integral is not so
straightforward as compared to the integral $I_2^{(d)}$. 
We will work out two-step procedure of functional reduction
allowing to express an integral $I_3^{(d)}$ that depends on
6 variables in terms of integrals depending on 3 variables.
\begin{center}
{\it Reduction of the integral $I_3^{(d)}$, step 1}
\end{center}

One of the solutions of the systems of equations (\ref{constrains}),
taken at $n=3$ and combined with
equations (\ref{I3_p0}), (\ref{I3_x_m}), leads to the desired relation
\begin{eqnarray}
&&I_3\left(m_1^2 ,m_2^2,m_3^2; s_{23},
s_{13},s_{12}\right)
\nonumber \\
&&
~~~~~~~~ =\kappa_{123}
I_3\left(r_{123} ,r_2,r_3; s_{23},r_3-r_{123},r_2-r_{123}\right)
\nonumber \\
&&
~~~~~~~~ +\kappa_{213}
I_3\left(r_{123},r_1,r_3; s_{13},r_3- r_{123},r_1-r_{123}\right) 
\nonumber \\
&&
~~~~~~~~ +\kappa_{312} I_3\left(r_{123},r_2,r_1; 
s_{12},r_1 -r_{123}, r_2-r_{123}\right),
\label{I3step1}
\end{eqnarray}
where 
\begin{eqnarray}
&&r_{123}=-\frac{\lambda_{123}}{g_{123}},~~~~r_i=m_i^2,
\nonumber \\
&&\kappa_{123}=\frac{\partial r_{123}}{\partial m_1^2},~~~
\kappa_{213}=\frac{\partial r_{123}}{\partial m_2^2},~~~
\kappa_{312}=\frac{\partial r_{123}}{\partial m_3^2},
\end{eqnarray}
and the determinants $\lambda_{123}$, $g_{123}$  are defined in Appendix.
Note that all integrals on the right side of equation 
(\ref{I3step1}) depend only on 4 variables.

This is not the only functional relation that reduces the number
of variables of the integral $I_3^{(d)}$.
We have discovered another functional relationship that reduces 
the number of variables by one 
\begin{eqnarray}
&&I_3\left(m_1^2 ,m_2^2,m_3^2; s_{23},
s_{13},s_{12}\right)
\nonumber \\
&&~~~~~~~ =\frac{s_{23}(s_{23}-s_{12}-s_{13})}{g_{123}}
I_3\left(m_0^2,m_2^2,m_3^2; s_{23},s_{123},s_{123}\right)
\nonumber \\
&&~~~~~~~ 
+\frac{s_{13}(s_{13}-s_{12}-s_{23})}{g_{123}}
I_3\left(m_1^2,m_0^2,m_3^2; s_{123},s_{13},s_{123}\right)
\nonumber \\
&&~~~~~~~ 
+\frac{s_{12}(s_{12}-s_{13}-s_{23})}{g_{123}}
I_3\left(m_1^2,m_2^2,m_0^2; s_{123},s_{123},s_{12}\right), 
\end{eqnarray}
where
\begin{equation}
m_0^2=-\frac{2\delta_3}{g_{123}},~~~~~~
s_{123}=-\frac{2s_{12}s_{13}s_{23}}{g_{123}},
\end{equation}
\begin{eqnarray}
&&
\delta_3 =- s_{12} s_{13} s_{23}
+
s_{23} (s_{12} + s_{13} - s_{23}) m_1^2
\nonumber \\
&&~~~~~~~~~~~~~~~~~~~~~
         + s_{13} (s_{12} - s_{13} + s_{23}) m_2^2
- s_{12} (s_{12} - s_{13} - s_{23}) m_3^2,~~~~~
\end{eqnarray}	 
and $g_{123}$ is defined in Appendix.
We have found many other functional relations that
reduce the number of variables, although not in all integrals at once.
However, as was shown in section 2, 
integrals without reducing the number 
of variables can be eliminated by combining various
functional relations (see Eq.~(\ref{I2fe_eqm})).
Derivation of this kind of functional relations
will be studied in more detail in a forthcoming publications.

\begin{center}
{\it Reduction of the integral $I_3^{(d)}$, step 2}
\end{center}
Now we proceed to the next step of the functional reduction. 
Applying relation (\ref{feI3massive}) to the first integral on the 
right side of Eq.~(\ref{I3step1}) and solving for the new 
variables $m_0^2$, $s_{0j}$, $x_k$ the corresponding
system of equations from (\ref{constrains}), combined with 
Eqs.~(\ref{I3_p0}), (\ref{I3_x_m}), leads to an equation
\begin{eqnarray}
\label{I3step2}
&&
I^{(d)}_3\left(r_{123},r_2,r_3; s_{23},r_3-r_{123},r_2-r_{123}\right)
\nonumber \\
&&~~~~~~~~~~=\kappa_{23}
I^{(d)}_3(r_{123},r_{23},r_3; r_3-r_{23},r_3-r_{123},r_{23}-r_{123})
\nonumber \\
&&~~~~~~~~~~+\kappa_{32}
I^{(d)}_3(r_{123},r_{23},r_2; r_2-r_{23},r_2-r_{123},r_{23}-r_{123}).
\end{eqnarray}
By appropriate change of variables, two more equations for reducing other
integrals in the right-hand side of Eq.~(\ref{I3step1}), can be obtained 
from Eq.~(\ref{I3step2}).

Combining  Eq.~(\ref{I3step1}), Eq.~(\ref{I3step2}) and two
equations that follow from Eq.~(\ref{I3step2}) by changing
variables, we get the final reduction formula for 
the integral $I_3^{(d)}$:
\begin{eqnarray}
\label{I3step1step2}
&&I_3^{(d)}(m_1^2,m_2^2,m_3^2; s_{23},s_{13},s_{12})
\nonumber \\
&&~~~~~~~~~~~~~~
  = \kappa_{123} \kappa_{23} I_3^{(d)}(r_{123},r_{23},r_3; r_3 - r_{23}, r_3 - r_{123}, r_{23} - r_{123})
\nonumber \\
&&~~~~~~~~~~~~~~
     + \kappa_{123}\kappa_{32} I_3^{(d)}(r_{123},r_{23},r_2; r_2 - r_{23}, r_2 - r_{123}, r_{23} - r_{123})
\nonumber \\
&&~~~~~~~~~~~~~~
+ \kappa_{213} \kappa_{31} I_3^{(d)}(r_{123},r_{13},r_1; r_1 - r_{13}, r_1 - r_{123}, r_{13} - r_{123})
\nonumber \\
&&~~~~~~~~~~~~~~
+ \kappa_{213} \kappa_{13} I_3^{(d)}(r_{123},r_{13},r_3; r_3 - r_{13}, r_3 - r_{123}, r_{13} - r_{123})
\nonumber \\
&&~~~~~~~~~~~~~~
+ \kappa_{312} \kappa_{12} I_3^{(d)}(r_{123},r_{12},r_2; r_2 - r_{12}, r_2 - r_{123}, r_{12} - r_{123})
\nonumber \\
&&~~~~~~~~~~~~~~
+ \kappa_{312} \kappa_{21} I_3^{(d)}(r_{123},r_{12},r_1; r_1 - r_{12}, r_1 - r_{123}, r_{12} - r_{123}).~~~~
\label{I3funcred}
\end{eqnarray}
This formula allows one to express the integral $I_3^{(d)}$,
which depends on 6 variables in terms of integrals depending only
on 3 variables.  

It is interesting to note that the replacement of masses and kinematic 
invariants on both sides of Eq.~(\ref{I3step1step2}) with arguments 
of the first integral $I_3^{(d)}$ on the right side
of this equation, i.e.
\begin{eqnarray}
&&m_1^2\rightarrow r_{123},~~~~ m_2\rightarrow r_{23},~~~~
m_3^2\rightarrow r_3, 
\nonumber \\
&&s_{23}\rightarrow r_3-r_{23},~~~~
s_{13}\rightarrow r_3-r_{123},~~~~s_{12}\rightarrow r_{23}-r_{123},
\label{I3_first_args}
\end{eqnarray}
leads to the following transformations of the coefficients and
arguments of integrals on the  right side of Eq.~(\ref{I3step1step2})
\begin{eqnarray}
&& r_{123} \rightarrow r_{123},~~~~\kappa_{123} \rightarrow 1,
~~~~\kappa_{213} \rightarrow 0,~~~~\kappa_{312} \rightarrow 0,
\nonumber \\
&& r_{23} \rightarrow r_{23},~~~~r_{13}\rightarrow r_{123},
~~~~r_{12}\rightarrow r_{123},
\nonumber \\
&&\kappa_{23}, \kappa_{13}, \kappa_{12} \rightarrow 1,~~~~
\kappa_{32}, \kappa_{31}, \kappa_{21} \rightarrow 0.
\end{eqnarray}
As expected, in Eq.~(\ref{I3funcred}), after these substitutions  
only the first term remains. Change of variables  (\ref{I3_first_args})
leads to a factorization of the determinants $\lambda$ and $g$
\begin{eqnarray}
&&\lambda_{123}=8r_{123}(r_{23}-r_3)(r_{123}-r_{23}),
~~~~g_{123}=-8(r_{23}-r_3)(r_{123}-r_{23}),
\nonumber \\
&&\lambda_{23}=-4r_{23}(r_{23}-r_3),~~~~g_{23}=-4(r_3-r_{23}),
\end{eqnarray}
and as follows from these relations,  $r_{123}$, $r_{23}$ remain   
invariant under  substitutions (\ref{I3_first_args}).

\subsection{Analytic results for integrals depending on the MNV}

Analytic result for the integral 
$I_3^{(d)}(r_{123},r_{23},r_3; r_3-r_{23},r_3-r_{123},r_{23}-r_{123})$ 
can be obtained, for example, 
 either by solving dimensional recurrence relation
 or by calculating the Feynman parameter integral.
 
 The dimensional recurrence relation for this integral reads
\begin{eqnarray}
&&(d-2)I_3^{(d+2)}(r_{123},r_{23},r_3; r_3-r_{23},
r_3-r_{123},r_{23}-r_{123})
=
\nonumber \\
&&~~~~~~~~~~~~-2 r_{123}
I_3^{(d)}(r_{123},r_{23},r_3; r_3-r_{23},
r_3-r_{123},r_{23}-r_{123})
\nonumber \\
&&~~~~~~~~~~~~~~
-I_2^{(d)}(r_{23},r_3; r_3-r_{23}).
\label{I3dimrec}
\end{eqnarray}
The solution of the dimensional recurrence relation (\ref{I3dimrec})  
was obtained by the method described in ref. \cite{Tarasov:2000sf}.
Assuming that
$\left|r_3/r_{123}\right|<1$, 
$\left|r_3/r_{23}\right|<1$, $\left|r_{23}/r_{123}\right|<1$,
we found the following result:
\begin{eqnarray}
&&I_3^{(d)}(r_{123},r_{23},r_3; r_3-r_{23},r_3-r_{123},r_{23}-r_{123})=
\nonumber \\
&&~~~
\frac{1}{\sin\frac{\pi d}{2}}\left\{
\frac{ r_{123}^{\frac{d-6}{2}}}{ \Gamma\left(\frac{d-2}{2}\right)}
C_3(x,y)
\right.
\nonumber \\
&&~~~+\frac{\pi^{\frac32} r_{23}^{\frac{d-4}{2}}}
{4r_{123}\Gamma\left(\frac{d-1}{2}\right) } 
\sqrt{\frac{r_{23}}{r_{23}-r_3}}~
\Fh21\Frr{1,\frac{d-2}{2}}{\frac{d-1}{2}}
\nonumber
\\
&&~~~\left.
-\frac{\pi r_3^{\frac{d-2}{2} }}
{4\Gamma\left(\frac{d}{2}\right) 
(r_{23}-r_3)r_{123}}\sqrt{1-\frac{r_3}{r_{23}}}
~F_1\left(\frac{d-2}{2},1,\frac12,\frac{d}{2};
\frac{r_3}{r_{123}},\frac{r_{3}}{r_{23}}\right)
\right\}, ~~~~
\label{I3_dimrec_solu}
\end{eqnarray}
where
\begin{eqnarray}
&&C_3(x,y)=
\frac{\pi x y^2}{4(x^2-y^2)^{\frac12}}
\ln\left(\frac{x-(x^2-y^2)^{\frac12}}
               {x+(x^2-y^2)^{\frac12}
               }\right),
\end{eqnarray}
and variables $x$, $y$ are defined as
\begin{equation}
x=\sqrt{\frac{r_{123}}{r_{123}-r_3}},~~~~~
y=\sqrt{\frac{r_{123}}{r_{123}-r_{23}}}.
\label{xy_C3}
\end{equation}
The function $C_3(x,y)$ was derived from a system of differential equations 
\begin{eqnarray}
&& x\frac{\partial C_3(x,y)}{\partial x}
+y\frac{\partial C_3(x,y)}{\partial y}=2C_3(x,y),
\nonumber \\
&&
(x^2-y^2)x\frac{\partial C_3(x,y)}{\partial x}=
-y^2C_3(x,y)-\frac12 \pi x^2 y^2.
\label{syst_for_c3}
\end{eqnarray}
This system was obtained from the system of differential
equations for the integral $I_3^{(d)}$.
We would like to notice the coefficient $1/\sin (\pi d/2)$ 
 in front of braces  which is   singular at $ d=4$. Since the integral 
 $I_3^{(4)}$ is finite, the terms in braces  at $d=4$ must  
cancel. This fact makes it possible to easily obtain
the hypergeometric Appell 
 function $F_1$ at $d=4$ as a combination of logarithms
\begin{eqnarray}
&&
F_1\left(1,1,\frac12,2;\frac{r_3}{r_{123}},\frac{r_3}{r_{23}}\right)
\nonumber \\
&&~~~~~~~~~~~~~~~~~~~~=
\frac{x^2\sqrt{1-y^2}}{1-x^2}
\left[\ln\left(\frac{1+\sqrt{1-y^2}}{1-\sqrt{1-y^2}}\right)
+\ln\left(\frac{x-\sqrt{x^2-y^2}}{x+\sqrt{x^2-y^2}}\right)
\right],~~~~~~~
\end{eqnarray}
where $x$,$y$ are defined in (\ref{xy_C3}).
This expression has been checked   numerically
to a precision of at least   200 decimal digits.

Another hypergeometric representation of the integral $I_3^{(d)}$ 
was  derived directly from the Feynman parameter integral
\begin{eqnarray}
&&I_3^{(d)}(r_{123},r_{23},r_3; r_3-r_{23},
r_3-r_{123},r_{23}-r_{123})
= \nonumber \\
&&~  -\Gamma\left(3-\frac{d}{2}\right)
\int_0^1\!\int_0^1\!\!x_1 \left[r_{123}-(r_{123}-r_{23})x_1^2
-(r_{23}-r_3)x_1^2x_2^2\right]^{\frac{d}{2}-3}\!dx_1dx_2 .~~~~~~~
\label{I3param}
\end{eqnarray}
Expanding the integrand in powers of variables
\begin{equation}
z_1=\frac{r_{123}-r_{23}}{r_{123}},~~~
z_2=\frac{r_{23}-r_{3}}{r_{123}},
\end{equation}
assuming that $|z_1|<1$, $|z_2|<1$ and integrating
over $x_1$, $x_2$, we then get 
\begin{eqnarray}
&&I_3^{(d)}(r_{123},r_{23},r_3; r_3-r_{23},r_3-r_{123},r_{23}-r_{123})
\nonumber \\
&& \nonumber \\
&&~~~~~~~~~~~~~~=  -\frac12 \Ga{3-\frac{d}{2}}
\sum_{n_1,n_2=0}^{\infty}\left(3-\frac{d}{2}\right)_{n_1+n_2}
\frac{(1)_{n_1+n_2}}{(2)_{n_1+n_2}}
\frac{\left(\frac{1}{2}\right)_{n_2}}{\left(\frac{3}{2}\right)_{n_2}}
\frac{z_1^{n_1}}{n_1!} \frac{z_2^{n_2}}{n_2!}.~~~~~~~~~~~~~~
\label{I3_via_T2}
\end{eqnarray}
Here $(a)_k=\Gamma(a+k)/\Gamma(a)$ is the so-called Pochhammer symbol.
The double series in Eq.~(\ref{I3_via_T2}) can be written 
\cite{Anastasiou:1999ui} as  the hypergeometric function $S_1$ 
\begin{eqnarray}
&&I_3^{(d)}(r_{123},r_{23},r_3; r_3-r_{23},r_3-r_{123},r_{23}-r_{123})
\nonumber \\
&& \nonumber \\
&& = -\frac{1}{2}\Gamma\left(3-\frac{d}{2}\right)r_{123}^{\frac{d}{2}-3}~
S_1\left(3-\frac{d}{2},1,\frac12,2,\frac32,\frac{r_{23}-r_3}{r_{123}}
,1-\frac{r_{23}}{r_{123}} \right).
\end{eqnarray}
The definition of the hypergeometric function $S_1$ is given in Appendix.
Using the formula for the analytic continuation of the  function $S_1$
presented in ref.~\cite{Anastasiou:1999ui}, the integral $I_3^{(d)}$
can be written in terms of hypergeometric  functions $_2F_1$ and $F_1$ : 
\begin{eqnarray}
&&I_3^{(d)}(r_{123},r_{23},r_3; r_3-r_{23},r_3-r_{123},r_{23}-r_{123})
= \frac{\Gamma\left(2-\frac{d}{2}\right)}
{2(r_{123}-r_{23})} r_{123}^{\frac{d}{2}-2}
\nonumber \\
&&~\times
\Biggl\{\!\Fh21\FRR{1,\frac12}{\frac32}
-\frac{r_{23}^{\frac{d}{2}-2}}
{ r_{123}^{\frac{d}{2}-2}}
F_1\left(\frac12,1,2-\frac{d}{2},\frac32; 
\frac{r_{23}-r_{3}}{r_{23}-r_{123}},1-\frac{r_{3}}{r_{23}}\right)\!\Biggr\}.~~~~~~
\label{I3viaS1}
\end{eqnarray} 
The  formula for the analytic continuation of the function $S_1$ 
is given in  Appendix (see Eq.~(\ref{S1_analyt_cont})).
Note that the results in terms of
the hypergeometric function $S_1$ for some $I_3^{(d)}$ integrals  
were presented in ref.~\cite{Anastasiou:1999ui}.

\section{Functional reduction of the 4-point integral $I_4^{(d)}$}

Now we proceed to formulate a functional reduction
procedure for the 4-point integral.
For $n=4$, the algebraic relation (\ref{usual_props}) reads
\begin{equation}
\frac{1}{D_1 D_2 D_3 D_4}= \frac{x_1}{D_0 D_2 D_3 D_4}
+\frac{x_2}{D_1 D_0 D_3 D_4}+\frac{x_3}{D_1 D_2 D_0 D_4}
+\frac{x_4}{D_1 D_2 D_3 D_0}.
\label{4point_relation}
\end{equation}
Equation (\ref{4point_relation}) holds if
\begin{equation}
p_0=x_1p_1+x_2p_2+x_3p_3+x_4p_4,
\end{equation}
and the parameters $m_0^2$, $x_j$ obey the following system
of equations
\begin{eqnarray}
&&x_1+x_2+x_3+x_4=1,\nonumber \\
&&x_1x_2s_{12}+x_1x_3s_{13}+x_1x_4s_{14}+x_2x_3s_{23}+
x_2x_4s_{24}+x_3x_4s_{34}
\nonumber \\
&&~~~~~~~~~~~~~~~~~~~~~~~~~~~~~~~~~
-x_1m_1^2-x_2m_2^2-x_3m_3^2-x_4m_4^2+m_0^2=0.~~~~~~~
\label{I4equsX}
\end{eqnarray}
Integrating equation (\ref{4point_relation})
over momentum $k_1$ yields
\begin{eqnarray}
&&I_4^{(d)}(m_1^2,m_2^2,m_3^2,m_4^2;
s_{12},s_{23},s_{34},s_{14},s_{24},s_{13})
\nonumber \\
&&~~~~~~~~~~~~~~~~~~~~=x_1
I_4^{(d)}(m_0^2,m_2^2,m_3^2,m_4^2;
s_{02},s_{23},s_{34},s_{04},s_{24},s_{03})
\nonumber \\
&&~~~~~~~~~~~~~~~~~~~~+x_{2}I_4^{(d)}(m_1^2,m_0^2,m_3^2,m_4^2;
s_{01},s_{03},s_{34},s_{14},s_{04},s_{13})
\nonumber \\
&&~~~~~~~~~~~~~~~~~~~~+x_{3}I_4^{(d)}(m_1^2,m_2^2,m_0^2,m_4^2;
s_{12},s_{02},s_{04},s_{14},s_{24},s_{01})
\nonumber \\
&&~~~~~~~~~~~~~~~~~~~~+x_4
I_4^{(d)}(m_1^2,m_2^2,m_3^2,m_0^2;
s_{12},s_{23},s_{03},s_{01},s_{02},s_{13}).
\label{box_func_equ}
\end{eqnarray}
This will be our initial equation for  deriving functional 
relations in all three steps of the reduction procedure.

\subsection{Functional reduction procedure}

In this subsection, we provide functional relations
for expressing the integral $I_4^{(d)}$ that depends
on 10 variables in terms of integrals $I_4^{(d)}$ depending
on 4 variables.

\begin{center}
{\it Reduction of the integral  $I_4^{(d)}$,  step 1}
\end{center}

Solving various systems of equations, formed from equations 
(\ref{I4equsX}) combined with equations (\ref{constrains}), 
taken at $n=4$, we obtained one solution
which leads to the functional relation that reduces
the number of variables by 3 in all integrals on the right side
of Eq.~(\ref{box_func_equ}). The functional relation 
corresponding to this solution reads
\begin{eqnarray}
&&I_4(m_1^2,m_2^2,m_3^2,m_4^2; s_{12},s_{23},s_{34},s_{14},s_{24},s_{13})
\nonumber \\
&& ~~=
\kappa_{1234}
I_4(r_{1234},r_2,r_3,r_4; r_2-r_{1234},s_{23},s_{34},r_4-r_{1234},s_{24},r_3-r_{1234})
\nonumber \\
&&~~
+\kappa_{2134}
I_4(r_{1234},r_1,r_3,r_4; r_1-r_{1234},s_{13},s_{34},r_4-r_{1234},s_{14},r_3-r_{1234})
\nonumber \\
&& ~~
+\kappa_{3124}
I_4(r_{1234},r_2,r_1,r_4; r_2-r_{1234},s_{12},s_{14},r_4-r_{1234},s_{24},r_1-r_{1234})
\nonumber \\
&& ~~
+\kappa_{4123}
I_4(r_{1234},r_2,r_3,r_1; r_2-r_{1234},s_{23},s_{13},r_1-r_{1234},s_{12},r_3-r_{1234}),~~~~~
\label{I4step1}
\end{eqnarray}
where
\begin{eqnarray}
&&r_{1234}=-\frac{\lambda_{1234}}{g_{1234}},~~~~~r_i=m_i^2,
\nonumber \\
&&\kappa_{1234}=\frac{\partial r_{1234}}{\partial m_1^2},
~~\kappa_{2134}=\frac{\partial r_{1234}}{\partial m_2^2},
~~\kappa_{3124}=\frac{\partial r_{1234}}{\partial m_3^2},
~~\kappa_{4123}=\frac{\partial r_{1234}}{\partial m_4^2}.~~~~~
\end{eqnarray}
In the next step, the integrals on the right side of equation (\ref{I4step1}) 
depending on 7 variables
will be expressed in terms of integrals depending on 5 variables.

\begin{center}
{\it Reduction of the integral $I_4^{(d)}$,  step 2}
\end{center}

Applying the formula (\ref{box_func_equ})  
to the first integral on the right side of Eq.~(\ref{I4step1})
and solving systems of equations formed from equations
(\ref{I4equsX}) combined with equations
(\ref{constrains}) given for the kinematics of this integral,
we found the following relation
\begin{eqnarray}
&&
I_4(r_{1234},r_2,r_3,r_4; r_2-r_{1234},s_{23},s_{34},r_4-r_{1234},s_{24},r_3-r_{1234})
\nonumber \\
&&=\kappa_{234}
I_4(r_{1234},r_{234},r_3,r_4; r_{234}-r_{1234},r_3-r_{234},s_{34},
r_4-r_{1234},r_4-r_{234},r_3-r_{1234})
\nonumber \\
&&+\kappa_{324}
I_4(r_{1234},r_{234},r_2,r_4; r_{234}-r_{1234},r_2-r_{234},s_{24},
r_4-r_{1234},r_4-r_{234},r_2-r_{1234})
\nonumber \\
&&+\kappa_{423}
I_4(r_{1234},r_{234},r_3,r_2; r_{234}-r_{1234},r_3-r_{234},s_{23},
r_2-r_{1234},r_2-r_{234},r_3-r_{1234}),
\nonumber \\
\label{I4step2}
\end{eqnarray}
where 
\begin{equation}
r_{234}=-\frac{\lambda_{234}}{g_{234}},
~~~~~~~\kappa_{234}=\frac{\partial r_{234}}{\partial m_2^2},
~~~~~~~\kappa_{324}=\frac{\partial r_{234}}{\partial m_3^2},
~~~~~~~\kappa_{423}=\frac{\partial r_{234}}{\partial m_4^2}.
\end{equation}
Similar expressions for all other integrals on the right side
of Eq.~(\ref{I4step1})  can be obtained from Eq.~(\ref{I4step2}) 
by  changing variables and coefficients appropriately.
 
After reducing  integrals depending on 7 variables 
to integrals depending on 5 variables,
the next step is to reduce the latter integrals to integrals
depending on 4 variables.

\begin{center}
{\it Reduction of the integral $I_4^{(d)}$,  step 3}
\end{center}
Applying our initial functional relation (\ref{box_func_equ})
to the first integral on the right side of Eq.~(\ref{I4step2})
and solving, the systems of equations, corresponding to this case,
we obtained several solutions.  One of these solutions 
leads to a two-term functional relation
\begin{eqnarray}
&&
I_4(r_{1234},r_{234},r_3,r_4; r_{234}-r_{1234},r_3-r_{234},s_{34},
r_4-r_{1234},r_4-r_{234},r_3-r_{1234})
\nonumber \\
&&=\kappa_{34}I_4(r_{1234},r_{234},r_{34},r_4;
\nonumber \\
&&~~~~~~~~~~~~~~~~~
r_{234}-r_{1234},r_{34}-r_{234},
r_4-r_{34},r_4-r_{1234},r_4-r_{234},r_{34}-r_{1234})
\nonumber \\
&&+\kappa_{43}
I_4(r_{1234},r_{234},r_{34},r_3; 
\nonumber \\
&&~~~~~~~~~~~~~~~~~
r_{234}-r_{1234},r_{34}-r_{234},
r_3-r_{34},r_3-r_{1234},r_3-r_{234},r_{34}-r_{1234}),
\nonumber \\
\label{I4step3}
\end{eqnarray}
where
\begin{equation}
r_{34}=-\frac{\lambda_{34}}{g_{34}},
~~~~~\kappa_{34}=\frac{\partial r_{34}}{\partial m_3^2},
~~~~~\kappa_{43}=\frac{\partial r_{34}}{\partial m_4^2}.
\end{equation}
Note that both integrals on the right side of Eq.~(\ref{I4step3})
depend only on 4 variables. Combining Eqs.~(\ref{I4step1}), (\ref{I4step2}),
(\ref{I4step3}) and all required relations 
that follow from these equations by changing variables
as mentioned  previously,
we obtain the final  functional reduction
formula for the integral $I_4^{(d)}$
\begin{eqnarray}
&&I_4(m_1^2,m_2^2,m_3^2,m_4^2; s_{12},s_{23},s_{34},s_{14},s_{24},s_{13})
\nonumber \\
&&=
\kappa_{1234} \kappa_{234} \kappa_{34} I_4^{(d)}(r_{1234}, r_{234}, r_{34}, r_4; 
\nonumber \\
&&~~~~~~~~~~
r_{234} - r_{1234}, r_{34} - r_{234},
    r_4 - r_{34}, r_4 - r_{1234}, r_4 - r_{234}, r_{34} - r_{1234})
\nonumber \\
&&
+\kappa_{1234} \kappa_{234} \kappa_{43} I_4^{(d)}(r_{1234}, r_{234}, r_{34}, r_3;  
\nonumber \\
&&~~~~~~~~~~~
r_{234} - r_{1234}, r_{34} - r_{234},
    r_3 - r_{34}, r_3 - r_{1234}, r_3 - r_{234}, r_{34} - r_{1234})
\nonumber \\
&&
+\kappa_{1234} \kappa_{324} \kappa_{24} I_4^{(d)}(r_{1234}, r_{234}, r_{24}, r_4;  
\nonumber 
\\
&&~~~~~~~~~~~
r_{234} - r_{1234}, r_{24} - r_{234},
    r_4 - r_{24}, r_4 - r_{1234}, r_4 - r_{234}, r_{24} - r_{1234})
\nonumber \\
&&
+\kappa_{1234} \kappa_{324} \kappa_{42} 
I_4^{(d)}(r_{1234}, r_{234}, r_{24}, r_2;  
\nonumber \\
&&~~~~~~~~~~~
r_{234} - r_{1234}, r_{24} - r_{234},
    r_2 - r_{24}, r_2 - r_{1234}, r_2 - r_{234}, r_{24} - r_{1234})
\nonumber 
\end{eqnarray}
\begin{eqnarray}
&&
+\kappa_{1234} \kappa_{423} \kappa_{23} I_4^{(d)}(r_{1234}, r_{234}, r_{23}, r_3; 
\nonumber \\
&&~~~~~~~~~~~
r_{234} - r_{1234}, r_{23} - r_{234},
    r_3 - r_{23}, r_3 - r_{1234}, r_3 - r_{234}, r_{23} - r_{1234})
\nonumber \\
&&
+\kappa_{1234} \kappa_{423} \kappa_{32} I_4^{(d)}(r_{1234}, r_{234}, r_{23}, r_2;  
\nonumber \\
&&~~~~~~~~~~~
r_{234} - r_{1234}, r_{23} - r_{234},
    r_2 - r_{23}, r_2 - r_{1234}, r_2 - r_{234}, r_{23} - r_{1234})
\nonumber \\
&&
+\kappa_{2134} \kappa_{134} \kappa_{34} I_4^{(d)}(r_{1234}, r_{134}, r_{34}, r_4;  
\nonumber \\
&&~~~~~~~~~~~
r_{134} - r_{1234}, r_{34} - r_{134},
    r_4 - r_{34}, r_4 - r_{1234}, r_4 - r_{134}, r_{34} - r_{1234})
\nonumber \\
&&
+\kappa_{2134} \kappa_{134} \kappa_{43} I_4^{(d)}(r_{1234}, r_{134}, r_{34}, r_3;  
\nonumber \\
&&~~~~~~~~~~~
r_{134} - r_{1234}, r_{34} - r_{134},
    r_3 - r_{34}, r_3 - r_{1234}, r_3 - r_{134}, r_{34} - r_{1234})
\nonumber \\
&&
+\kappa_{2134} \kappa_{314} \kappa_{14} I_4^{(d)}(r_{1234}, r_{134}, r_{14}, r_4; 
\nonumber \\
&&~~~~~~~~~~~
r_{134} - r_{1234}, r_{14} - r_{134},
    r_4 - r_{14}, r_4 - r_{1234}, r_4 - r_{134}, r_{14} - r_{1234})
\nonumber \\
&&
+\kappa_{2134} \kappa_{314} \kappa_{41} I_4^{(d)}(r_{1234}, r_{134}, r_{14}, r_1; 
\nonumber \\
&&~~~~~~~~~~~
r_{134} - r_{1234}, r_{14} - r_{134},
    r_1 - r_{14}, r_1 - r_{1234}, r_1 - r_{134}, r_{14} - r_{1234})
\nonumber \\
&&
+\kappa_{2134} \kappa_{413} \kappa_{13} I_4^{(d)}(r_{1234}, r_{134}, r_{13}, r_3;
\nonumber \\
&&~~~~~~~~~~~
r_{134} - r_{1234}, r_{13} - r_{134},
    r_3 - r_{13}, r_3 - r_{1234}, r_3 - r_{134}, r_{13} - r_{1234})
\nonumber \\
&&
+\kappa_{2134} \kappa_{413} \kappa_{31} I_4^{(d)}(r_{1234}, r_{134}, r_{13}, r_1; 
\nonumber \\
&&~~~~~~~~~~~
r_{134} - r_{1234}, r_{13} - r_{134},
    r_1 - r_{13}, r_1 - r_{1234}, r_1 - r_{134}, r_{13} - r_{1234})
\nonumber \\
&&
+\kappa_{3124} \kappa_{124} \kappa_{24} I_4^{(d)}(r_{1234}, r_{124}, r_{24}, r_4; 
\nonumber \\
&&~~~~~~~~~~~
r_{124} - r_{1234}, r_{24} - r_{124},
    r_4 - r_{24}, r_4 - r_{1234}, r_4 - r_{124}, r_{24} - r_{1234})
\nonumber \\
&&
+\kappa_{3124} \kappa_{124} \kappa_{42} I_4^{(d)}(r_{1234}, r_{124}, r_{24}, r_2;  
\nonumber \\
&&~~~~~~~~~~~
r_{124} - r_{1234}, r_{24} - r_{124},
    r_2 - r_{24}, r_2 - r_{1234}, r_2 - r_{124}, r_{24} - r_{1234})
\nonumber \\
&&
+\kappa_{3124} \kappa_{214} \kappa_{14} I_4^{(d)}(r_{1234}, r_{124}, r_{14}, r_4; 
\nonumber \\
&&~~~~~~~~~~~
r_{124} - r_{1234}, r_{14} - r_{124},
    r_4 - r_{14}, r_4 - r_{1234}, r_4 - r_{124}, r_{14} - r_{1234})
\nonumber \\
&&
+\kappa_{3124} \kappa_{214} \kappa_{41} I_4^{(d)}(r_{1234}, r_{124}, r_{14}, r_1; 
\nonumber \\
&&~~~~~~~~~~~
r_{124} - r_{1234}, r_{14} - r_{124},
    r_1 - r_{14}, r_1 - r_{1234}, r_1 - r_{124}, r_{14} - r_{1234})
\nonumber \\
&&
+\kappa_{3124} \kappa_{412} \kappa_{12} I_4^{(d)}(r_{1234}, r_{124}, r_{12}, r_2; 
\nonumber \\
&&~~~~~~~~~~~
r_{124} - r_{1234}, r_{12} - r_{124},
    r_2 - r_{12}, r_2 - r_{1234}, r_2 - r_{124}, r_{12} - r_{1234})
\nonumber \\
&&
+\kappa_{3124} \kappa_{412} \kappa_{21} I_4^{(d)}(r_{1234}, r_{124}, r_{12}, r_1; 
\nonumber \\
&&~~~~~~~~~~~
r_{124} - r_{1234}, r_{12} - r_{124},
    r_1 - r_{12}, r_1 - r_{1234}, r_1 - r_{124}, r_{12} - r_{1234})
\nonumber \\
&&
+\kappa_{4123} \kappa_{123} \kappa_{23} I_4^{(d)}(r_{1234}, r_{123}, r_{23}, r_3; 
\nonumber \\
&&~~~~~~~~~~~
r_{123} - r_{1234}, r_{23} - r_{123},
    r_3 - r_{23}, r_3 - r_{1234}, r_3 - r_{123}, r_{23} - r_{1234})
\nonumber \\
&&
+\kappa_{4123} \kappa_{123} \kappa_{32} I_4^{(d)}(r_{1234}, r_{123}, r_{23}, r_2; 
\nonumber \\
&&~~~~~~~~~~~
r_{123} - r_{1234}, r_{23} - r_{123},
    r_2 - r_{23}, r_2 - r_{1234}, r_2 - r_{123}, r_{23} - r_{1234})
\nonumber \\
&&
+\kappa_{4123} \kappa_{213} \kappa_{13} I_4^{(d)}(r_{1234}, r_{123}, r_{13}, r_3; 
\nonumber \\
&&~~~~~~~~~~~
r_{123} - r_{1234}, r_{13} - r_{123},
    r_3 - r_{13}, r_3 - r_{1234}, r_3 - r_{123}, r_{13} - r_{1234})
\nonumber 
\end{eqnarray}
\begin{eqnarray}
&&
+\kappa_{4123} \kappa_{213} \kappa_{31} I_4^{(d)}(r_{1234}, r_{123}, r_{13}, r_1; 
\nonumber \\
&&~~~~~~~~~~~
r_{123} - r_{1234}, r_{13} - r_{123},
    r_1 - r_{13}, r_1 - r_{1234}, r_1 - r_{123}, r_{13} - r_{1234})
\nonumber \\
&&
+\kappa_{4123} \kappa_{312} \kappa_{12} I_4^{(d)}(r_{1234}, r_{123}, r_{12}, r_2;  
\nonumber \\
&&~~~~~~~~~~~
r_{123} - r_{1234}, r_{12} - r_{123},
    r_2 - r_{12}, r_2 - r_{1234}, r_2 - r_{123}, r_{12} - r_{1234})
\nonumber \\
&&
+\kappa_{4123} \kappa_{312} \kappa_{21} I_4^{(d)}(r_{1234}, r_{123}, r_{12}, r_1; 
\nonumber \\
&&~~~~~~~~
r_{123} - r_{1234}, r_{12} - r_{123},
    r_1 - r_{12}, r_1 - r_{1234}, r_1 - r_{123}, r_{12} - r_{1234}).~~~
\label{I4_final}
\end{eqnarray} 
This formula allows us to express an integral $I_4^{(d)}$, that depends
on 15 variables, as a linear combination of 24 integrals each of which 
depends only on 4 variables.

We conclude this subsection by noting an interesting property of this
equation.  By replacing the masses and kinematic invariants 
in Eq.~(\ref{I4_final})  with corresponding arguments of the first
integral on the right side of this equation
\begin{eqnarray}
&&m_1^2 \rightarrow r_{1234},~~~~m_2^2\rightarrow r_{234},
~~~~m_3^2 \rightarrow r_{34},~~~~m_4^2\rightarrow r_4,
\nonumber \\
&&s_{12}\rightarrow r_{234}-r_{1234},~~~~
s_{23}\rightarrow r_{34}-r_{234},~~~~
s_{34}\rightarrow r_{4}-r_{34},~~~~
\nonumber\\
&&
s_{14}\rightarrow r_{4}-r_{1234},~~~~
s_{24}\rightarrow r_{4}-r_{234},~~~~
s_{13}\rightarrow r_{34}-r_{1234},~~~~
\label{I4invariance}
\end{eqnarray}
we get  the following transformations of the 
coefficients and arguments of  integrals in this equation
\begin{eqnarray}
&&r_{1234}\rightarrow r_{1234},~~~
r_{234}\rightarrow r_{234},~~~
r_{34}\rightarrow r_{34},~~~
\nonumber \\
&&
\kappa_{1234}, \kappa_{234},\kappa_{34} \rightarrow 1,~~~~
\kappa_{2134},\kappa_{3124},\kappa_{4123} \rightarrow 0.
\end{eqnarray}
As a result of these substitutions, only the first integral remains 
on the right side of the equation. 
Note that the  transformations (\ref{I4invariance}) lead
to a factorization of the determinants $\lambda$ and $g$ such as
\begin{eqnarray}
&&
\lambda_{1234} = 16 r_{1234} (r_{34} - r_4)(r_{34} - r_{234})
(r_{1234} - r_{234}),
\nonumber \\
&&
g_{1234} =  -16 (r_{34} - r_4) (r_{34} - r_{234})
(r_{1234} - r_{234}).
\end{eqnarray}
Similar factorization holds for lower order determinants.
We observed  analogous factorization of determinants
appearing in the final functional relations for integrals $I_5^{(d)}$
and  $I_6^{(d)}$.

\subsection{Analytic results for integrals depending on the MNV}

Analytic expression for an integral 
\begin{equation}
I_4^{(d)}(m^2_1,m^2_2,m^2_3,m^2_4; m^2_2-m^2_1,m^2_3-m^2_2,
m^2_4-m^2_3,m^2_4-m^2_1,m^2_4-m^2_2,m^2_3-m^2_1),
\end{equation}
that depends on the MNV, can be obtained, for example,  by solving 
dimensional recurrence relation or by evaluating the Feynman parameter
integral.

The dimensional recurrence relation for this integral reads
\begin{eqnarray}
&&
(d-3)   I_4^{(d+2)}(r_{1234},r_{234},r_{34},r_4; 
\nonumber \\
&&~~~~~~~
   r_{234}-r_{1234},r_{34}-r_{234},r_4-r_{34},r_4-r_{1234},
   r_4-r_{234},r_{34}-r_{1234})=
\nonumber \\  
&& \nonumber \\
&&    -2r_{1234} I_4^{(d)}(r_{1234},r_{234},r_{34},r_4; 
\nonumber \\
&&~~~~~~~
   r_{234}-r_{1234},r_{34}-r_{234},r_4-r_{34},
   r_4-r_{1234},r_4-r_{234},r_{34}-r_{1234})
\nonumber \\  
&& \nonumber \\
&&~~~~~~~~~~~~~~
- I_3^{(d)}(r_{234}, r_{34}, r_4; r_4 - r_{34}, r_4 - r_{234}, r_{34} - r_{234} ).
\end{eqnarray}
To solve this equation, we used for the integral $I_3^{(d)}$
an analytic result given in Eq.~(\ref{I3_dimrec_solu}).
Applying the method descibed in ref. \cite{Tarasov:2000sf}, we get 
\begin{eqnarray}
&&I_4^{(d)}(r_{1234},r_{234},r_{34},r_4;
\nonumber \\
&&~~~~~~~~~~~~~~~
r_{234}-r_{1234},r_{34}-r_{234},r_4-r_{34},r_4-r_{1234},
   r_4-r_{234},r_{34}-r_{1234} )=
\nonumber \\
&&\frac{1}{\sin{\frac{\pi d}{2}}}\Biggl\{
\frac{r_{1234}^{\frac{d}{2}-4}}
{\Gamma\left(\frac{d-3}{2}\right)}
C_4(x,y,z)
\nonumber \\
&& ~~~~ +\frac{\pi r_{234}^{\frac{d}{2}-2}
\arctan{\frac{(r_{34}-r_4)^{\frac12}}{(r_{234}-r_{34})^{\frac12}  }}}
{4r_{1234}(r_{34}-r_4)^{\frac12}(r_{234}-r_{34})^{\frac12}
\Gamma\left(\frac{d-2}{2}\right)}
\Fh21\FOO{1,\frac{d-3}{2}}{\frac{d-2}{2}}
\nonumber \\
&&~~~~ -\frac{\pi^{\frac32}r_{34}^{\frac{d}{2}-2}}
{8r_{1234}r_{234}\Gamma\left(\frac{d-1}{2}\right)}
\left(\frac{r_{34}}{r_{34}-r_4}\right)^{\frac12}
\left(\frac{r_{234}}{r_{234}-r_{34}}\right)^{\frac12}
\nonumber \\
&&~~~~~~~~~~~~~~~~~~\times
F_1\left(\frac{d-3}{2},1,\frac12,\frac{d-1}{2};
\frac{r_{34}}{r_{1234}},\frac{r_{34}}{r_{234}}\right)
\nonumber \\
&&~~~~
 +\frac{\pi r_4^{\frac{d}{2}-1}}{8r_{1234}(r_{34}-r_4)(r_{234}-r_4)
 \Gamma\left(\frac{d}{2}\right)}
 \nonumber \\
 &&~~~~~~~~~~~ \times 
 F_S\left(\frac{d-3}{2},1,1,1,1,\frac12,\frac{d}{2},
 \frac{d}{2},\frac{d}{2};\frac{r_4}{r_{1234}},\frac{r_4}{r_4-r_{234}},
 \frac{r_4}{r_4-r_{34}}\right)   \Biggr\},
\label{I4solution_dr}
\end{eqnarray}   
where
\begin{eqnarray}
C_4(x,y,z) =
\frac{\pi^{\frac32}~x y^2 z^2}
{8 (x^2-y^2)^{\frac12} (z^2-y^2)^{\frac12}}
\ln \left[\frac{x z+y^2-(z^2-y^2)^{\frac12}(x^2-y^2)^{\frac12} }
{x z+y^2+(z^2-y^2)^{\frac12}(x^2-y^2)^{\frac12} }
\right],~~~
\label{C4}
\end{eqnarray}
and the definition of the hypergeometric Lauricella-Saran function 
$F_S$ is given in Appendix (see Eq.~(\ref{FSseries})).
Here the variables $x,y,z$ are defined as
\begin{equation}
x=\sqrt{\frac{r_{1234}}{r_{1234}-r_4}},~~~~
y=\sqrt{\frac{r_{1234}}{r_{1234}-r_{34}}},~~~~
z=\sqrt{\frac{r_{1234}}{r_{1234}-r_{234}}}.
\label{xyz}
\end{equation}
The function $C_4(x,y,z)$ was obtained by solving the system 
of differential equations
\begin{eqnarray}
&& x\frac{\partial C_4(x,y,z)}{\partial x}
+y\frac{\partial C_4(x,y,z)}{\partial y}
+z\frac{\partial C_4(x,y,z)}{\partial z}
=3C_4(x,y,z),
\nonumber \\
&&
\frac{\partial C_4(x,y,z)}{\partial x}=
-\frac{y^2}{x(x^2-y^2)}C_4(x,y,z)
-\frac{\pi^{\frac32} x y^2 z^2}{4(x^2-y^2)(x+z)}
\nonumber \\
&& \nonumber \\
&&
\frac{\partial C_4(x,y,z)}{\partial z}=
\frac{ (2y^2-z^2)}{z(y^2-z^2)}C_4(x,y,z)
+\frac{\pi^{\frac32}~ x y^2  z^2}{4(y^2-z^2)(x+z)}.
\label{C4difsyst}
\end{eqnarray}
This system was derived from a system of differential equations
for the integral $I_4^{(d)}$. The constant of integration that
occurs in solving  a system of differential equations 
(\ref{C4difsyst})
was determined by comparing the asymptotic behavior of Eq.~(\ref{I4solution_dr}) 
with the asymptotic behavior of the integral $I_4^{(d)}$
at $r_{1234} \rightarrow \infty$.

Note that in calculations of Feynman integrals, the Lauricella-Saran 
function  $F_S$  was first discovered
when calculating the one-loop box integral  \cite{Fleischer:2003rm}. 
In ref.~\cite{Fleischer:2003rm}, it was shown 
that this function can be represented by a one-fold integral
\begin{eqnarray}
&&F_S\left( \frac{d-3}{2},1,1; 1,1,\frac12;
\frac{d}{2},\frac{d}{2},\frac{d}{2};
x,y,z \right) \nonumber \\
&&~~~~~~~~~~~~~~~~
=\frac{\Gamma \left(\frac{d}{2}\right)~(y-z)^{-\frac12}  }
{\Gamma\left(\frac{d-3}{2}\right) \Gamma\left(\frac32\right)}
 \int_0^1 \frac{\arcsin \sqrt{\frac{(y-z)t}{1-tz}}
 (1-t)^{\frac{d-5}{2}}}{(1-x+tx) \sqrt{1-ty}} ~dt .
\label{intrepFS}
\end{eqnarray}

We note that, similarly to the integral $I_3^{(d)}$,
there is a factor $1/\sin (\pi d/2)$ 
 in front of braces in Eq.~(\ref{I4solution_dr})  which is 
 singular at $d=4$. Since the integral  $I_4^{(4)}$ is finite, the terms
 in braces must  cancel  at $d=4$. This fact makes it possible 
 to easily obtain
the hypergeometric function $F_S$ at $d=4$ as a combination of logarithms
\begin{eqnarray}
&&
F_S\left(\frac12,1,1,1,1,\frac12,2,
 2,2;\frac{r_4}{r_{1234}},\frac{r_4}{r_4-r_{234}},
 \frac{r_4}{r_4-r_{34}}\right)
 \nonumber \\
&&~~=
\frac{(x^2-y^2)^{\frac12}(x^2-z^2)}
     {(z^2-y^2)^{\frac12}(1-x^2)x}
\left\{
\ln \left[\frac{x z+y^2-(z^2-y^2)^{\frac12}(x^2-y^2)^{\frac12} }
{x z+y^2+(z^2-y^2)^{\frac12}(x^2-y^2)^{\frac12} }\right]
\right.
\nonumber\\
&&~~
\left.
-\ln \left[\frac{  x(y^2-z^2)^{\frac12}-z(y^2-x^2)^{\frac12}}
     {  x(y^2-z^2)^{\frac12}+z(y^2-x^2)^{\frac12}}\right]
+\ln \left[\frac{(z^2-y^2)^{\frac12}-(1-y^2)^{\frac12}}
                 {(z^2-y^2)^{\frac12}+(1-y^2)^{\frac12}}\right]
\right\},
\end{eqnarray}
where $x,y$ and $z$ are defined in Eq.~(\ref{xyz}).
This expression has been checked   numerically
to a precision of at least   200 decimal digits.
Complications in analytic calculation of a periodic function that
appears in  solving the recurrence relation 
for the one-loop box integral were discussed in ref.~\cite{Phan:2018cnz}.

The integral $I_4^{(d)}$ depending on the MNV can be represented as a triple
hypergeometric series. 
Such a representation can be derived from the 
Feynman parameter integral
\begin{eqnarray}
&&I_4^{(d)}(r_{1234},r_{234},r_{34},r_4; 
r_{234}-r_{1234},r_{34}-r_{234},r_3-r_{34},r_4-r_{234},r_{34}-
r_{1234})
\nonumber \\
&& \nonumber \\
&&~~~~~~~~~~~~~~~~~~~~~~~ = 
\Gamma\left(4-\frac{d}{2}\right)\int_0^1\int_0^1\int_0^1
~x_1^2  x_2~ h_4^{\frac{d}{2}-4} dx_1dx_2dx_3,
\label{I4_param_rep}
\end{eqnarray}
where
\begin{equation}
h_4=r_{1234}
-(r_{1234}-r_{234})x_1^2
-(r_{234}-r_{34})x_1^2 x_2^2
-(r_{34}-r_4)x_1^2 x_2^2x_3^2,
\label{h4}
\end{equation}
Expanding the integrand of (\ref{I4_param_rep}) 
with respect to the three variables
\begin{equation}
z_1=1-\frac{r_{234}}{r_{1234}},~~~~z_2=\frac{r_{234}-r_{34}}{r_{1234}},
~~~~z_3=\frac{r_{34}-r_4}{r_{1234}},
\end{equation}
assuming that $|z_1|<1$, $|z_2|<1$, $|z_3|<1$ and integrating 
over $x_1$, $x_2$, $x_3$ term by term, we obtain
the series representation 
\begin{eqnarray}
&&I_4^{(d)}(r_{1234},r_{234},r_{34},r_4; 
r_{234}-r_{1234},r_{34}-r_{234},r_3-r_{34},r_4-r_{234},r_{34}-
r_{1234})
\nonumber \\
&& \nonumber \\
&&~~~~~~~
=\frac16 \Gamma\left(4-\frac{d}{2}\right)r_{1234}^{\frac{d}{2}-4}
\sum_{n_1,n_2,n_3=0}^{\infty}
\left(4-\frac{d}{2}\right)_{n_1+n_2+n_3}
\nonumber \\
&&~~~~~~~~~~~~~~~~~~~~~~~~~~~~~~~~~\times
\frac{\left(\frac32\right)_{n_1+n_2+n_3}}
{\left(\frac52\right)_{n_1+n_2+n_3}}
\frac{\left( 1 \right)_{n_2+n_3}}{\left( 2\right)_{n_2+n_3}}
\frac{\left(\frac12\right)_{n_3}}{\left(\frac32\right)_{n_3}}
\frac{z_1^{n_1} z_2^{n_2}z_3^{n_3}}{n_1!n_2!n_3!}.
\label{T3series}
\end{eqnarray}
This series can be written as a Kamp\'e de F\'eriet function
\cite{Appell:1926:FHPH}.
The analytic continuation of this series can be expressed in terms 
of the hypergeometric function  $_2F_1$, the Appell functions
$F_1$, $F_3$ and various hypergeometric Lauricella-Saran functions. 
The formula (\ref{I4solution_dr}) is an example of such a representation.
Other examples can be  found in refs. \cite{Bytev:2013gva}, 
\cite{Davydychev:2017bbl}.
The relationship between the hypergeometric Lauricella-Saran functions
 $F_S$ and $F_N$ is given in Appendix (see Eq.~(\ref{FS_via_FN})).

\section{Functional reduction of the 5-point integral $I_5^{(d)}$ }
In this section we describe the functional reduction of the 5-point 
integral. At $n=5$ the algebraic relation that follows from 
the Eq.~(\ref{usual_props}) reads
\begin{eqnarray}
&&\frac{1}{D_1 D_2 D_3 D_4 D_5}=
\frac{x_1}{D_0 D_2 D_3 D_4D_5}
\nonumber \\
&&~~~
+\frac{x_2}{D_1 D_0 D_3 D_4 D_5}+\frac{x_3}{D_1 D_2 D_0 D_4 D_5}
+\frac{x_4}{D_1 D_2 D_3 D_0D_5}
+\frac{x_5}{D_1 D_2 D_3 D_4D_0}.~~~~~~~
\label{5prop_relation}
\end{eqnarray}
Equation (\ref{5prop_relation}) is valid if
\begin{equation}
p_0=x_1p_1+x_2p_2+x_3p_3+x_4p_4 +x_5p_5,
\label{I5p0}
\end{equation}
and the parameters $m_0^2$, $x_j$ obey the following system
of equations
\begin{eqnarray}
&&x_1+x_2+x_3+x_4+x_5=1,\nonumber \\
&&x_1x_2s_{12}+x_1x_3s_{13}+x_1x_4s_{14}+x_1x_5s_{15}
+x_2x_3s_{23}+x_2x_4s_{24}
\nonumber \\
&&~~~~~~+x_2x_5s_{25}+x_3x_4s_{34}+x_3x_5s_{35}+x_4x_5s_{45}
\nonumber \\
&&~~~~~~
-x_1m_1^2-x_2m_2^2-x_3m_3^2-x_4m_4^2-x_5m_5^2+m_0^2=0.~~~~~~~
\label{I5mm0}
\end{eqnarray}
Integrating equation (\ref{5prop_relation}) over $k_1$ yields a 
functional relation
\begin{eqnarray}
&&I^{(d)}_5(m^2_1,m^2_2,m^2_3,m^2_4,m^2_5; 
      s_{12},s_{23},s_{34},s_{45},s_{15},s_{13},s_{14},s_{24},s_{25},s_{35})
\nonumber \\
&&
=
x_1 I^{(d)}_5(m^2_{0},m^2_2,m_3^2,m_4^2,m^2_5; 
      s_{02},s_{23},s_{34},s_{45},s_{05},s_{03},s_{04},s_{24},s_{25},s_{35})
\nonumber \\
&&+
x_2 I^{(d)}_5(m_1^2,m^2_0,m_3^2,m_4^2,m^2_5; 
      s_{01},s_{03},s_{34},s_{45},s_{15},s_{13},s_{14},s_{04},s_{05},s_{35})
\nonumber \\
&&+
x_3 I^{(d)}_5(m_1^2,m^2_2,m_0^2,m_4^2,m^2_5; 
      s_{12},s_{02},s_{04},s_{45},s_{15},s_{01},s_{14},s_{24},s_{25},s_{05})
\nonumber \\      
&&+
x_4 I^{(d)}_5(m_1^2,m^2_2,m_3^2,m_0^2,m^2_5; 
      s_{12},s_{23},s_{03},s_{05},s_{15},s_{13},s_{01},s_{02},s_{25},s_{35})
\nonumber \\      
&&
+x_5 I^{(d)}_5(m_1^2,m^2_2,m_3^2,m_4^2,m^2_0; 
      s_{12},s_{23},s_{34},s_{04},s_{01},s_{13},s_{14},s_{24},s_{02},s_{03}).
\label{penta_func_equ}
\end{eqnarray}
This equation will be our initial equation in deriving
functional relations in all stages
of the reduction procedure.

\subsection{Functional reduction procedure}
In this subsection, we describe four steps of the 
functional reduction procedure, that will allow us to represent the
integral $I_5^{(d)}$, that depends on $15$ variables 
in terms of integrals depending on $5$ variables.

\begin{center}
{\it Reduction of the integral $I_5^{(d)}$,  step 1}
\end{center}
Systems of equations  that were formed from a set of equations 
(\ref{constrains}) and  Eqs.~(\ref{I5p0}), (\ref{I5mm0}), have
many solutions. One of these solutions leads to the following relation
\begin{eqnarray}
&&I^{(d)}_5(m^2_1,m^2_2,m_3^2,m_4^2,m^2_5;
      s_{12},s_{23},s_{34},s_{45},s_{15},s_{13},s_{14},s_{24},s_{25},s_{35})
\nonumber \\
&&
=
\kappa_{12345} I^{(d)}_5(r_{12345},r_2,r_3,r_4,r_5;
\nonumber \\
&&~~~~~~
     r_2-r_{12345},s_{23},s_{34},s_{45},r_5-r_{12345},
     r_3-r_{12345}, r_4-r_{12345},s_{24},s_{25},s_{35})
\nonumber \\
&&+
\kappa_{21345} I^{(d)}_5(r_1,r_{12345},r_3,r_4,r_5;
\nonumber \\
&&~~~~~~
      r_1-r_{12345},r_3-r_{12345} ,s_{34},s_{45},s_{15},s_{13},s_{14},r_4-r_{12345},
      r_5-r_{12345},s_{35})
\nonumber \\
&&+
\kappa_{31245} I^{(d)}_5(r_1,r_2,r_{12345},r_4,r_5;
\nonumber \\
&&~~~~~~
      s_{12},r_2-r_{12345},r_4-r_{12345},s_{45},s_{15},
      r_1-r_{12345},s_{14},s_{24},s_{25},r_5-r_{12345})
\nonumber \\      
&&+
\kappa_{41235} I^{(d)}_5(r_1,r_2,r_3,r_{12345},r_5;
\nonumber \\
&&~~~~~~
      s_{12},s_{23},r_3-r_{12345},r_5-r_{12345},s_{15},s_{13},
      r_1-r_{12345},r_2-r_{12345},s_{25},s_{35})
\nonumber \\      
&&
+\kappa_{51234} I^{(d)}_5(r_1,r_2,r_3,r_4,r_{12345}; 
\nonumber \\
&&~~~~~~
      s_{12},s_{23},s_{34},r_4-r_{12345},r_1-r_{12345},s_{13},s_{14},s_{24},
      r_2-r_{12345},r_3-r_{12345}),
\nonumber \\      
\label{I5step1}
\end{eqnarray}
where
\begin{eqnarray}
&&
r_{12345}=-\frac{\lambda_{12345}}{g_{12345}},
~~~\kappa_{12345}=\frac{\partial r_{12345}}{\partial m_1^2},
~~~\kappa_{21345}=\frac{\partial r_{12345}}{\partial m_2^2},~~~r_i=m_i^2,
\nonumber \\
&&
\kappa_{31245}=~\frac{\partial r_{12345}}{\partial m_3^2},
~~~\kappa_{41235}=\frac{\partial r_{12345}}{\partial m_4^2},
~~~\kappa_{51234}=\frac{\partial r_{12345}}{\partial m_5^2}.
\label{mm0x5}
\nonumber 
\end{eqnarray}
This equation is the first step in the functional reduction
procedure, expressing an integral depending on 15 variables
as a linear combination of integrals depending on 11 variables.

\begin{center}
{\it Reduction of the integral $I_5^{(d)}$,  step 2}
\end{center}

By setting the  masses and kinematic variables in  equation (\ref{penta_func_equ})
to be equal to the arguments of the first integral in the right-hand side
of the equation (\ref{I5step1}) and solving corresponding sets of systems of
equations, we found
\begin{eqnarray}
&&I^{(d)}_5(r_{12345},r_2,r_3,r_4,r_5;
\nonumber \\
&&~~~~~~
     r_2-r_{12345},s_{23},s_{34},s_{45},r_5-r_{12345},
     r_3-r_{12345}, r_4-r_{12345},s_{24},s_{25},s_{35})
\nonumber \\
&&= \kappa_{2345}
I^{(d)}_5(r_{12345},r_{2345},r_3,r_4,r_5; r_{2345}-r_{12345}, r_3-r_{2345},
s_{34},
\nonumber \\
&&~~~~~~~~~~~
     s_{45},r_5-r_{12345},
     r_3-r_{12345}, r_4-r_{12345},r_4-r_{2345},r_5-r_{2345},s_{35})
\nonumber \\
&&+\kappa_{3245}
I^{(d)}_5(r_{12345},r_2,r_{2345},r_4,r_5;
     r_2-r_{12345},r_2-r_{2345},r_4-r_{2345},
\nonumber \\
&&~~~~~~~~~~~
     s_{45},r_5-r_{12345},
     r_{2345}-r_{12345}, r_4-r_{12345},s_{24},s_{25},r_5-r_{2345})
\nonumber \\     
&&+\kappa_{4235}
I^{(d)}_5(r_{12345},r_2,r_3,r_{2345},r_5;
     r_2-r_{12345},s_{23},r_3-r_{2345},
r_5-r_{2345},
     \nonumber \\
&&~~~~~~~~~~~
     r_5-r_{12345},
     r_3-r_{12345}, r_{2345}-r_{12345},r_2-r_{2345},s_{25},s_{35})
\nonumber \\
&&+\kappa_{5234}
I^{(d)}_5(r_{12345},r_2,r_3,r_4,r_{2345};
     r_2-r_{12345},s_{23},s_{34},r_4-r_{2345},
\nonumber \\
&&~~~~~~~~~~~
     r_{2345}-r_{12345},
     r_3-r_{12345}, r_4-r_{12345},s_{24},r_2-r_{2345},r_3-r_{2345}),
\nonumber \\  
\label{I5step2}
\end{eqnarray}
where
\begin{eqnarray}
&&
r_{2345}=-\frac{\lambda_{2345}}{g_{2345}},
\nonumber \\
&&
\kappa_{2345}=\frac{\partial r_{2345}}{\partial m_2^2},
~~~\kappa_{3245}=\frac{\partial r_{2345}}{\partial m_3^2},
~~~\kappa_{4235}=\frac{\partial r_{2345}}{\partial m_4^2},
~~~\kappa_{5234}=\frac{\partial r_{2345}}{\partial m_5^2}.
\end{eqnarray}
Eq.~(\ref{I5step2}) allows one to express the first integral
on the right-hand side of equation (\ref{I5step1}), which depends
on 11 variables, in terms of integrals depending on 8 variables.
Similar equations that reduce the number of variables by 3
for other integrals in the right-hand side of equation (\ref{I5step1})
can be obtained by  the  proper change of arguments and coefficients 
in equation (\ref{I5step2}).

\begin{center}
{\it Reduction of the integral $I_5^{(d)}$,  step 3}
\end{center}

Integrals depending on 8 variables can be expressed in terms of
integrals depending on 6 variables.
Similar to the previous steps, we apply relation (\ref{penta_func_equ})
to the first integral on the right-hand side of equation (\ref{I5step2}),
and solving the corresponding sets of systems of equations, we get
\begin{eqnarray}
&&I^{(d)}_5(r_{12345},r_{2345},r_3,r_4,r_5; r_{2345}-r_{12345}, r_3-r_{2345},
\nonumber \\
&&~~~~~~~~~~~
     s_{34},s_{45},r_5-r_{12345},
     r_3-r_{12345}, r_4-r_{12345},r_4-r_{2345},r_5-r_{2345},s_{35})
\nonumber \\
&&= \kappa_{345}
I^{(d)}_5(r_{12345},r_{2345},r_{345},r_4,r_5; r_{2345}-r_{12345}, r_{345}-r_{2345},
r_4-r_{345}, 
\nonumber \\
&&~~~~~~~~~~~
     s_{45},r_5-r_{12345},
     r_{345}-r_{12345}, r_4-r_{12345},r_4-r_{2345},r_5-r_{2345},r_5- r_{345})
\nonumber \\
&&+\kappa_{435} I^{(d)}_5(r_{12345},r_{2345},r_3,r_{345},r_5;
r_{2345}-r_{12345}, r_3-r_{2345},
r_3-r_{345},
\nonumber \\
&&~~~~~~~~~~~
     r_5-r_{345},r_5-r_{12345},
     r_3-r_{12345}, r_{345}-r_{12345},r_{345}-r_{2345},r_5-r_{2345},s_{35})
 \nonumber\\
&&+\kappa_{534}
I^{(d)}_5(r_{12345},r_{2345},r_3,r_4,r_{345};
r_{2345}-r_{12345}, r_3-r_{2345},
s_{34},r_4-r_{345},
\nonumber \\
&&~~~~~~~~~~~
     r_{345}-r_{12345},
     r_3-r_{12345}, r_4-r_{12345},r_4-r_{2345},r_{345}-r_{2345},r_3-r_{345}),
\nonumber \\
\label{I5step3}
 \end{eqnarray} 
where
 \begin{eqnarray*}
&&
r_{345}=-\frac{\lambda_{345}}{g_{345}},
~~~\kappa_{345}=\frac{\partial r_{345}}{\partial m_3^2},
~~~\kappa_{435}=\frac{\partial r_{345}}{\partial m_4^2},
~~~\kappa_{534}=\frac{\partial r_{345}}{\partial m_5^2}.
\end{eqnarray*}
The functional relation (\ref{I5step3}) reduces an integral depending
on 8 variables to a linear combination of integrals depending on 6 variables.
From equation (\ref{I5step3}) one can obtain similar equations for
reducing 
all the other integrals in the right side of Eq.~(\ref{I5step2}).

\begin{center}
{\it Reduction of the integral $I_5^{(d)}$,  step 4}
\end{center}

Now we proceed to derive the last set of  reduction equations. To do
this, we apply relation (\ref{penta_func_equ}) to the first integral 
on the right side of (\ref{I5step3}), solve systems of equations
consisting of equations (\ref{constrains}), (\ref{I5p0}), (\ref{I5mm0})
and find 
\begin{eqnarray}
&&I^{(d)}_5(r_{12345},r_{2345},r_{345},r_4,r_5; r_{2345}-r_{12345}, r_{345}-r_{2345},
r_4-r_{345},s_{45},
\nonumber \\
&&~~~~~~~~~~~
     r_5-r_{12345},
     r_{345}-r_{12345}, r_4-r_{12345},r_4-r_{2345},r_5-r_{2345},r_5- r_{345})
\nonumber \\
&&=\kappa_{45}
I^{(d)}_5(r_{12345},r_{2345},r_{345},r_{45},r_5; r_{2345}-r_{12345}, r_{345}-r_{2345},
r_{45}-r_{345},r_5 -r_{45},
\nonumber \\
&&~~~~~~~~~~~
     r_5-r_{12345},
     r_{345}-r_{12345}, r_{45}-r_{12345},r_{45}-r_{2345},r_5-r_{2345},r_5- r_{345})
\nonumber \\
&&+\kappa_{54} I^{(d)}_5(r_{12345},r_{2345},r_{345},r_4,r_{45};
r_{2345}-r_{12345}, r_{345}-r_{2345},
r_4-r_{345},r_4-r_{45},
\nonumber \\
&&~~~~~~~~~~~
     r_{45}-r_{12345},
     r_{345}-r_{12345}, r_4-r_{12345},r_4-r_{2345},r_{45}-r_{2345},r_{45}- r_{345}),
\nonumber \\
\label{I5step4}
\end{eqnarray}
where
\begin{eqnarray*}
&&
r_{45}=-\frac{\lambda_{45}}{g_{45}},
~~~~~~\kappa_{45}=\frac{\partial r_{45}}{\partial m_4^2},
~~~~~~\kappa_{54}=\frac{\partial r_{45}}{\partial m_5^2}.
\end{eqnarray*}
Again, functional relations for reducing  all  other integrals 
on the right side of Eq.~(\ref{I5step3}) can be obtained from 
(\ref{I5step4}) by the proper change of variables.

By combining  equations (\ref{I5step1}), (\ref{I5step2}), 
 (\ref{I5step3}), (\ref{I5step4}), and all required relations 
that follow from these equations by changing variables
and coefficients as mentioned  previously,
we obtained a formula that will allow us to express 
the integral $I_5^{(d)}$, depending on 15 variables, as a linear 
combination of 120 integrals, each of which  depends
on only  5 variables. All resulting  integrals in this combination
depend on the MNV and have the form
\begin{eqnarray}
&&
I_5^{(d)}(m^2_i,m^2_j,m^2_k,m^2_l,m^2_r;
~m^2_j-m^2_i,m^2_k-m^2_j,m^2_l-m^2_k,m^2_r-m^2_l,m^2_r-m^2_i,
\nonumber \\
&&~~~~~~~~~~~~~~~~~~~~~~~~~~
     m^2_k-m^2_i,m^2_l-m^2_i,m^2_l-m^2_j,m^2_r-m^2_j,m^2_r-m^2_k),
\end{eqnarray}     
where $m^2_i$, $m^2_j$,$m^2_k$,$m^2_l$,$m^2_r$ 
are  ratios of polynomials in  masses and  kinematic invariants.
Coefficients in front of these integrals are also
ratios of polynomials in masses  and  kinematic invariants.

The final reduction formula for the integral $I_5^{(d)}$ 
is too lengthy to present in the manuscript.
Instead, we provide these formulae in electronic form
in ancillary files distributed with this article.

\subsection{Dimensional
recurrence relation and series representation }

The dimensional recurrence relation for the integral $I_5^{(d)}$ 
depending on the MNV reads
\begin{eqnarray}
&&(d-4)I_5^{(d+2)}(r_{12345},r_{2345},r_{345},r_{45},r_5; 
r_{2345}-r_{12345},r_{345}-r_{2345}, r_{45}-r_{345},
\nonumber \\
&&~~~~~
          r_5-r_{45},r_5-r_{12345},r_{345}-r_{12345},r_{45}-r_{12345},
 	  r_{45}-r_{2345},r_5-r_{2345},r_5-r_{345})=
\nonumber \\	  
&& \nonumber \\
&&-2r_{12345}I_5^{(d)}(r_{12345},r_{2345},r_{345},r_{45},r_5; 
r_{2345}-r_{12345},r_{345}-r_{2345},r_{45}-r_{345},
\nonumber \\
&&~~~~~
           r_5-r_{45},r_5-r_{12345},r_{345}-r_{12345},r_{45}-r_{12345},
 	  r_{45}-r_{2345},r_5-r_{2345},r_5-r_{345})
\nonumber \\
&& \nonumber \\
&&        
    -I_4^{( d)}(
    r_{2345}, r_{345}, r_{45}, r_5;
\nonumber\\
&&~~~~~~~~~
    r_{345} - r_{2345}, r_{45} - r_{345}, r_5 - r_{45}, r_5 - r_{2345}, r_5 -
     r_{345}, r_{45} - r_{2345}).~~~~
\label{I5dimrec}     
\end{eqnarray}
Notice that the inhomogeneous part of this equation 
consists of only one term - the integral $I_4^{(d)}$ which
also depends on the MNV.
Solution of the dimensional recurrence relation for the integral
$I_5^{(d)}$ is a bit cumbersome but straightforward. We will present
the result and details of the derivation in a separate publication.

At $d=4$ the term with the integral $I_5^{(d+2)}$ in Eq.~(\ref{I5dimrec})  
drops out, and we get a simple relation
\begin{eqnarray}
&&2r_{12345}I_5^{(4)}(r_{12345},r_{2345},r_{345},r_{45},r_5; 
r_{2345}-r_{12345},r_{345}-r_{2345},r_{45}-r_{345},
\nonumber \\
&&~~~~~
           r_5-r_{45},r_5-r_{12345},r_{345}-r_{12345},r_{45}-r_{12345},
 	  r_{45}-r_{2345},r_5-r_{2345},r_5-r_{345})
\nonumber \\
&& \nonumber \\
&& =       
    -I_4^{(4)}(
    r_{2345}, r_{345}, r_{45}, r_5;
\nonumber\\
&&~~~~~~~~~
    r_{345} - r_{2345}, r_{45} - r_{345}, r_5 - r_{45}, r_5 - r_{2345}, r_5 -
     r_{345}, r_{45} - r_{2345}).~~~~
\end{eqnarray}

Now we proceed to derive multiple series representation for the 
integral $I_5^{(d)}$ depending on the MNV. To do this, we will use 
the following Feynman parameter integral representation 
\begin{eqnarray}
&&I_5^{(d)}(r_{12345},r_{2345},r_{345},r_{45},r_5; 
r_{2345}-r_{12345},r_{345}-r_{2345},r_{45}-r_{345},r_5-r_{45},
r_5-r_{12345},
\nonumber \\
&&~~~~~~
r_{345}-r_{12345},r_{45}-r_{12345},r_{45}-r_{2345} , r_5-r_{2345},r_5-r_{345})
=\nonumber \\
&&~~~~~~~~~-\Gamma\left(5-\frac{d}{2}\right)
\int_0^1\int_0^1\int_0^1 \int_0^1 x_1^3x_2^2x_3  h_5^{\frac{d}{2}-5}
dx_1dx_2dx_3dx_4,
\label{I5paramrep}
\end{eqnarray}
where
\begin{eqnarray}
&&
h_5=r_{12345}-(r_{12345}-r_{2345})x_1^2
\nonumber \\
&&~~~~~~
-(r_{2345}-r_{345}) x_1^2x_2^2
-(r_{345}-r_{45}) x_1^2 x_2^2 x_3^2-(r_{45}-r_5) x_1^2 x_2^2 x_3^2 x_4^2.~~
\end{eqnarray}

Multiple series representation of the  integral $I_5^{(d)}$ can be
derived 
in a similar manner as it was done for the integrals $I_3^{(d)}$,
$I_4^{(d)}$.
Expanding the integrand in Eq.~(\ref{I5paramrep}) with respect to
the four variables
\begin{equation}
z_1=\frac{r_{12345}-r_{2345}}{r_{12345}},
~~~
z_2=\frac{r_{2345}-r_{345}}{r_{12345}},
~~~
z_3=\frac{r_{345}-r_{45}}{r_{12345}},
~~~
z_4=\frac{r_{45}-r_{5}}{r_{12345}},
\end{equation}
assuming that all $|z_i|<1$, and then integrating
 term by term over $x_1$,...,$x_4$, we get
\begin{eqnarray}
&&I_5^{(d)}(r_{12345},r_{2345},r_{345},r_{45},r_5;
r_{2345}-r_{12345},r_{345}-r_{2345}, r_{45}-r_{345},
r_5-r_{45},
\nonumber \\
&&~~~~~
          r_5-r_{12345},r_{345}-r_{12345},r_{45}-r_{12345},
 	  r_{45}-r_{2345},r_5-r_{2345},r_5-r_{345})
\nonumber \\	
&& \nonumber \\
&&~~~~=
-\frac{1}{24}\Ga{5-\frac{d}{2}}
\sum_{n_1,n_2,n_3,n_4=0}^{\infty}\left(5-\frac{d}{2}\right)_{n_1+n_2+n_3+n_4}
\nonumber \\
&&~~~~~~~~~~~~~\times
\frac{(2)_{n_1+n_2+n_3+n_4}}{(3)_{n_1+n_2+n_3+n_4}}
\frac{\left(\frac32\right)_{n_2+n_3+n_4}}{\left(\frac52\right)_{n_2+n_3+n_4}}
\frac{(1)_{n_3+n_4}}{(2)_{n_3+n_4}}
\frac{\left(\frac12\right)_{n_4}}{\left(\frac32\right)_{n_4}}
\frac{z_1^{n_1}}{n_1!}\frac{z_2^{n_2}}{n_2!}\frac{z_3^{n_3}}{n_3!}
\frac{z_4^{n_4}}{n_4!}.
\label{I5viaT4}
\end{eqnarray}
Note the similarity of the summand of this series to summands
of the series given in Eqs.~(\ref{I3_via_T2}), (\ref{T3series}).

\section{Functional reduction of the 6-point integral $I_6^{(d)}$}
Now we proceed to derive  functional relations for reducing
the 6-point integral.
At $n=6$ the algebraic relation (\ref{usual_props}) reads
\begin{eqnarray}
&&\frac{1}{D_1 D_2 D_3 D_4 D_5D_6}=
\frac{x_1}{D_0 D_2 D_3 D_4D_5D_6}
+\frac{x_2}{D_1 D_0 D_3 D_4 D_5D_6}
+\frac{x_3}{D_1 D_2 D_0 D_4 D_5D_6}
\nonumber \\
&&~~~
+\frac{x_4}{D_1 D_2 D_3 D_0D_5D_6}
+\frac{x_5}{D_1 D_2 D_3 D_4D_0D_6}
+\frac{x_6}{D_1 D_2 D_3 D_4D_5D_0}.
\label{6prop_relation}
\end{eqnarray}
Equation (\ref{6prop_relation}) is valid if
\begin{equation}
p_0=x_1p_1+x_2p_2+x_3p_3+x_4p_4 +x_5p_5+x_6p_6,
\label{I6p0}
\end{equation}
and the parameters $m_0^2$, $x_j$ obey the following system
of equations
\begin{eqnarray}
&&x_1+x_2+x_3+x_4+x_5+x_6=1,\nonumber \\
&&x_1x_2s_{12}+x_1x_3s_{13}+x_1x_4s_{14}+x_1x_5s_{15}+x_1x_6s_{16}
+x_2x_3s_{23}+x_2x_4s_{24}
\nonumber \\
&&~~
+x_2x_5s_{25}+x_2x_6s_{26}
+x_3x_4s_{34}
+x_3x_5s_{35}+x_3x_6s_{36}+x_4x_5s_{45}
+x_4x_6s_{46}
\nonumber \\
&&~~+x_5x_6s_{56}
-x_1m_1^2-x_2m_2^2-x_3m_3^2-x_4m_4^2-x_5m_5^2
-x_6m_6^2+m_0^2=0.~~~~~~~
\label{I6mm0}
\end{eqnarray}
Integrating  both sides of Eq.~(\ref{6prop_relation}) over $k_1$
yields a functional relation
\begin{eqnarray}
&&I^{(d)}_6(m_1^2,m_2^2,m_3^2,m_4^2,m_5^2,m_6^2;
      s_{12},s_{23},s_{34},s_{45},s_{56},s_{16},
\nonumber\\
&&~~~~~~~~~~~~~~~~~~~~~~~~~~~~~~~~~~~~~
      s_{13},s_{14},s_{15},s_{24},s_{25},s_{26},s_{35},s_{36},s_{46})
\nonumber \\
&&= x_1
I^{(d)}_6(m_0^2,m_2^2,m_3^2,m_4^2,m_5^2,m_6^2;
      s_{02},s_{23},s_{34},s_{45},s_{56},s_{06},
\nonumber\\
&&~~~~~~~~~~~~~~~~~~~~~~~~~~~~~~~~~~~~~
      s_{03},s_{04},s_{05},s_{24},s_{25},s_{26},s_{35},s_{36},s_{46})
\nonumber \\
&&+x_2
I^{(d)}_6(m_1^2,m_0^2,m_3^2,m_4^2,m_5^2,m_6^2;
      s_{01},s_{03},s_{34},s_{45},s_{56},s_{16},
\nonumber\\
&&~~~~~~~~~~~~~~~~~~~~~~~~~~~~~~~~~~~~~
      s_{13},s_{14},s_{15},s_{04},s_{05},s_{06},s_{35},s_{36},s_{46})
\nonumber \\
&&+x_3
I^{(d)}_6(m_1^2,m_2^2,m_0^2,m_4^2,m_5^2,m_6^2;
      s_{12},s_{02},s_{04},s_{45},s_{56},s_{16},
\nonumber\\
&&~~~~~~~~~~~~~~~~~~~~~~~~~~~~~~~~~~~~~
      s_{01},s_{14},s_{15},s_{24},s_{25},s_{26},s_{05},s_{06},s_{46})
\nonumber \\
&&+x_4
I^{(d)}_6(m_1^2,m_2^2,m_3^2,m_0^2,m_5^2,m_6^2;
      s_{12},s_{23},s_{03},s_{05},s_{56},s_{16},
\nonumber\\
&&~~~~~~~~~~~~~~~~~~~~~~~~~~~~~~~~~~~~~
      s_{13},s_{01},s_{15},s_{02},s_{25},s_{26},s_{35},s_{36},s_{06})
\nonumber \\
&&+x_5
I^{(d)}_6(m_1^2,m_2^2,m_3^2,m_4^2,m_0^2,m_6^2;
      s_{12},s_{23},s_{34},s_{04},s_{06},s_{16},
\nonumber\\
&&~~~~~~~~~~~~~~~~~~~~~~~~~~~~~~~~~~~~~
      s_{13},s_{14},s_{01},s_{24},s_{02},s_{26},s_{03},s_{36},s_{46})
\nonumber \\
&&+x_6
I^{(d)}_6(m_1^2,m_2^2,m_3^2,m_4^2,m_5^2,m_0^2;
      s_{12},s_{23},s_{34},s_{45},s_{05},s_{01},
\nonumber\\
&&~~~~~~~~~~~~~~~~~~~~~~~~~~~~~~~~~~~~~
      s_{13},s_{14},s_{15},s_{24},s_{25},s_{02},s_{35},s_{03},s_{04}).
\label{I6_ini_fe}
\end{eqnarray}
This equation will be used in all steps of functional reduction
of the integral $I_6^{(d)}$.
Derivation of the reduction formulae  is completely analogous
to that of the integrals $I_2^{(d)}$,...,$I_5^{(d)}$.
 
\subsection{Functional reduction procedure}
In this subsection we will describe five steps of
functional reduction procedure allowing us to represent the
integral $I_6^{(d)}$ depending on $21$ variables 
in terms of integrals depending on $6$ variables.

\begin{center}
{\it Reduction of the integral $I_6^{(d)}$,  step 1}
\end{center}

In the first step we
formed various systems of equations by combining Eqs.~(\ref{constrains}),
taken at $n=6$, and Eqs.~(\ref{I6p0}), (\ref{I6mm0}) and solved
these systems for $x_j$, $m_0^2$. Many solutions have been found.
One of these solutions leads to a functional relation reducing 
5 variables simultaneously for all integrals
in the right-hand side of Eq.~(\ref{I6_ini_fe}). 
The functional relation corresponding to this solution reads
\begin{eqnarray}
&&I^{(d)}_6(m_1^2,m_2^2,m_3^2,m_4^2,m_5^2,m_6^2;
      s_{12},s_{23},s_{34},s_{45},s_{56},s_{16},
\nonumber \\
&&~~~~~~~~~~~~~~~~~~~~~~~~~~~~~
      s_{13},s_{14},s_{15},
      s_{24},s_{25},s_{26}, s_{35},s_{36},s_{46} )
\nonumber \\
&& \nonumber \\
&&=\kappa_{123456}
I^{(d)}_6(r_{123456},r_2,r_3,r_4,r_5,r_6;
      r_2- r_{123456},s_{23},s_{34},s_{45},s_{56},
      r_6 -r_{123456},
      \nonumber \\
&&~~~~~~~~~~~~     
      r_3-r_{123456},r_4-r_{123456},r_5 - r_{123456} ,
      s_{24},s_{25},s_{26}, s_{35},s_{36},s_{46} )
\nonumber \\
&& \nonumber \\
&&+\kappa_{213456}
I^{(d)}_6(r_1,r_{123456},r_3,r_4,r_5,r_6;
      r_1-r_{123456},r_3-r_{123456},s_{34},s_{45},s_{56},s_{16},
  \nonumber \\
&&~~~~~~~~~~~~~
      s_{13},s_{14},s_{15},
     r_4-r_{123456}, r_5- r_{123456} ,r_6-r_{123456}, s_{35},s_{36},s_{46} )
\nonumber \\
&& \nonumber \\
&&+\kappa_{312456}
I^{(d)}_6(r_1,r_2,  r_{123456},r_4,r_5,r_6;
      s_{12},r_2- r_{123456} ,r_4-r_{123456},s_{45},s_{56},s_{16},
\nonumber \\
&&~~~~~~~~~~~~~~
      r_1 -r_{123456},s_{14},s_{15},
      s_{24},s_{25},s_{26}, r_5-r_{123456}, r_6 -r_{123456},s_{46} )
\nonumber 
\end{eqnarray}
\begin{eqnarray}
&&+\kappa_{412356}
I^{(d)}_6(r_1,r_2,r_3,r_{123456},r_5,r_6;
      s_{12},s_{23},r_3-r_{123456},r_5-r_{123456},s_{56},
\nonumber \\
&&~~~~~~~~~~~~~~
      s_{16},
      s_{13},r_1- r_{123456},s_{15},
      r_2-r_{123456} ,s_{25},s_{26}, s_{35},s_{36},r_6-r_{123456} )
\nonumber \\
\nonumber \\
&&+\kappa_{512346} I^{(d)}_6(r_1,r_2,r_3,r_4,r_{123456},r_6;
      s_{12},s_{23},s_{34},r_4- r_{123456},r_6 -r_{123456},
 \nonumber \\
&&~~~~~~~~~~~~~~
      s_{16},
      s_{13},s_{14},r_1 - r_{123456},
      s_{24},r_2 -r_{123456},s_{26},r_3-r_{123456},s_{36},s_{46} )
\nonumber \\
\nonumber \\
&&+\kappa_{612345} I^{(d)}_6(r_1,r_2,r_3,r_4,r_5,r_{123456};
      s_{12},s_{23},s_{34},s_{45},r_5 - r_{123456},
      r_1-r_{123456},
\nonumber \\
&&~~~~~~~~~~~~~~
      s_{13},s_{14},s_{15},
      s_{24},s_{25},r_2 -r_{123456}, s_{35},r_3-r_{123456},
      r_4 -r_{123456} ),
\nonumber \\
\label{I6step1}
      \end{eqnarray}
where
\begin{eqnarray}
&&
r_{123456}=-\frac{\lambda_{123456}}{g_{123456}},~
\kappa_{123456}=\frac{\partial r_{123456}}{\partial m_1^2},~
\kappa_{213456}=\frac{\partial r_{123456}}{\partial m_2^2},~
\kappa_{213456}=\frac{\partial r_{123456}}{\partial m_3^2},
\nonumber \\
&&
~~~~
\kappa_{213456}=\frac{\partial r_{123456}}{\partial m_4^2},~~
\kappa_{213456}=\frac{\partial r_{123456}}{\partial m_5^2},~~
\kappa_{612345}=\frac{\partial r_{123456}}{\partial m_6^2},~~r_i=m_i^2.
\end{eqnarray}
There are several other solutions of the systems of equations
(\ref{constrains}), (\ref{I6p0}), (\ref{I6mm0})  allowing us to
reduce the number of variables simultaneously for all integrals
in the right-hand side of Eq.~(\ref{I6_ini_fe}) but the number
of variables reducible by functional relations corresponding to
these solutions were less than 5.
We obtained also many solutions leading to the functional relations
reducing the number of variables but
not for all integrals simultaneously. Some of these solutions
depend on square roots of polynomials in kinematic variables 
and masses. 

\begin{center}
{\it Reduction of the integral $I_6^{(d)}$,  step 2}
\end{center}

In the second step of the reduction we take arguments of the first
integral in the right-hand side of equation (\ref{I6step1}) and
substitute them into the initial functional equation (\ref{I6_ini_fe}).
By solving
systems of equations composed of  equations  (\ref{constrains}) and
equations (\ref{I6p0}), (\ref{I6mm0}), 
for the new unknowns $m_0^2$, $s_{0j}$, $x_j$,  we 
found a solution allowing us to reduce four variables
simultaneously for all integrals. This solution
leads the following reduction formula
\begin{eqnarray}
&&I^{(d)}_6(r_{123456},r_2,r_3,r_4,r_5,r_6;
      r_2- r_{123456},s_{23},s_{34},s_{45},s_{56},
      r_6 -r_{123456},
      \nonumber \\
&&~~~~~~~~~~~~     
      r_3-r_{123456},r_4-r_{123456},r_5 - r_{123456} ,
      s_{24},s_{25},s_{26}, s_{35},s_{36},s_{46} )
\nonumber \\
&& \nonumber \\
&&=\kappa_{23456}
I^{(d)}_6(r_{123456},r_{23456},r_3,r_4,r_5,r_6;
      r_{23456}- r_{123456},r_3-r_{23456},s_{34},s_{45},s_{56},
      r_6 -r_{123456},
      \nonumber \\
&&~~~~~~~    
      r_3-r_{123456},r_4-r_{123456},r_5 - r_{123456} ,
      r_4-r_{23456},r_5-r_{23456},r_6-r_{23456}, s_{35},s_{36},s_{46} )
\nonumber \\
&& \nonumber \\
&&+\kappa_{32456}I^{(d)}_6(r_{123456},r_2,r_{23456},r_4,r_5,r_6;
      r_2- r_{123456},r_2 - r_{23456},r_4-r_{23456},s_{45},s_{56},
      r_6 -r_{123456},
      \nonumber \\
&&~~~~~~~~~~~~     
      r_{23456}-r_{123456},r_4-r_{123456},r_5 - r_{123456} ,
      s_{24},s_{25},s_{26}, r_5-r_{23456},r_6-r_{23456},s_{46} )
\nonumber \\
&& \nonumber\\
&&+\kappa_{42356}I^{(d)}_6(r_{123456},r_2,r_3, r_{23456},r_5,r_6;
      r_2- r_{123456},s_{23},r_3 -r_{23456},r_5-r_{23456},s_{56},
      r_6 -r_{123456},
      \nonumber \\
&&~~~~~~~~~~~~     
      r_3-r_{123456},r_{23456}-r_{123456},r_5 - r_{123456} ,
      r_2-r_{23456},s_{25},s_{26}, s_{35},s_{36},r_6-r_{23456} )
\nonumber \\
&& \nonumber \\
&&+\kappa_{52346}
I^{(d)}_6(r_{123456},r_2,r_3,r_4,r_{23456},r_6;
      r_2- r_{123456},s_{23},s_{34},r_4-r_{23456},
      r_6-r_{23456},
      r_6 -r_{123456},
      \nonumber \\
&&~~~~~~~~~~~~     
      r_3-r_{123456},r_4-r_{123456},r_{23456} - r_{123456} ,
      s_{24},r_2 -r_{23456},s_{26},r_3-r_{23456},s_{36},s_{46} )
\nonumber \\
&& \nonumber \\
&&+\kappa_{62345}I^{(d)}_6(r_{123456},r_2,r_3,r_4,r_5,r_{23456};
      r_2- r_{123456},s_{23},s_{34},s_{45},r_5-r_{23456},
      r_{23456} -r_{123456},
      \nonumber \\
&&~~~~~~~~~~~~     
      r_3-r_{123456},r_4-r_{123456},r_5 - r_{123456} ,
      s_{24},s_{25},r_2 -r_{23456}, s_{35},
      r_3-r_{23456},r_4-r_{23456} ),
\nonumber \\
\label{I6step2}
\end{eqnarray}
where
\begin{eqnarray}
&&r_{23456}=-\frac{\lambda_{23456}}{g_{23456}},
~~~\kappa_{23456}=\frac{\partial r_{23456}}{\partial m_2^2},
~~~\kappa_{32456}=\frac{\partial r_{23456}}{\partial m_3^2},
\nonumber \\
&&
\kappa_{42356}=\frac{\partial r_{23456}}{\partial m_4^2},
~~~\kappa_{52346}=\frac{\partial r_{23456}}{\partial m_5^2},
~~~\kappa_{62345}=\frac{\partial r_{23456}}{\partial m_6^2}.
\end{eqnarray}
Similar considerations apply to other integrals from the right-hand 
side of the equation (\ref{I6step1}).
Hence, with the aid of the functional relation (\ref{I6step2}),
the  original integral depending on 16 variables will be reduced to
a combination of integrals depending on 12 variables. 

\begin{center}
{\it Reduction of the integral $I_6^{(d)}$,  step 3}
\end{center}

In the third step, integrals depending on 12 variables were reduced
to integrals depending on 9 variables. 
We substitute arguments of the first integral
in the right side of equation (\ref{I6step2}) into our  
initial equation (\ref{I6_ini_fe}) and solve 
systems of equations composed of equations (\ref{constrains}) 
and (\ref{I6p0}), (\ref{I6mm0}) for the new unknowns $x_j$, $m_0^2$.

One of the obtained solutions yields  the required reduction formula
\begin{eqnarray*}
&&
I^{(d)}_6(r_{123456},r_{23456},r_3,r_4,r_5,r_6;
      r_{23456}- r_{123456},r_3-r_{23456},s_{34},s_{45},s_{56},
      r_6 -r_{123456},
      \nonumber \\
&&~~~~~~~    
      r_3-r_{123456},r_4-r_{123456},r_5 - r_{123456} ,
      r_4-r_{23456},r_5-r_{23456},r_6-r_{23456}, s_{35},s_{36},s_{46} )
\nonumber \\
&& \nonumber \\
&&=
\kappa_{3456}
I^{(d)}_6(r_{123456},r_{23456}, r_{3456},r_4,r_5,r_6;
      r_{23456}- r_{123456},r_{3456}-r_{23456},r_4-r_{3456},      
      \nonumber \\
&&~~~~~~~s_{45},s_{56},r_6 -r_{123456},    
      r_{3456}-r_{123456},r_4-r_{123456},r_5 - r_{123456} ,
      r_4-r_{23456},
\nonumber \\
&&~~~~~~~~r_5-r_{23456},
      r_6-r_{23456}, 
      r_5-r_{3456},r_6-r_{3456},s_{46} )
\nonumber \\
\end{eqnarray*}
\begin{eqnarray}
&&+\kappa_{4356}
I^{(d)}_6(r_{123456},r_{23456},r_3,r_{3456},r_5,r_6;
      r_{23456}- r_{123456},r_3-r_{23456},
      r_3-r_{3456},
\nonumber \\
&&~~~~~~~
      r_5-r_{3456},s_{56},
      r_6 -r_{123456},
      r_3-r_{123456},r_{3456}-r_{123456},
      \nonumber \\
&&~~~~~~~    
      r_5 - r_{123456} ,
      r_{3456}-r_{23456},r_5-r_{23456},r_6-r_{23456}, s_{35},s_{36},
      r_6-r_{3456} )
\nonumber \\
&& \nonumber \\
&&+\kappa_{5346}
I^{(d)}_6(r_{123456},r_{23456},r_3,r_4,r_{3456},r_6;
      r_{23456}- r_{123456},r_3-r_{23456},s_{34},
      \nonumber \\
&&~~~~~~~    
      r_4-r_{3456},r_6-r_{3456},
      r_6 -r_{123456},
      r_3-r_{123456},r_4-r_{123456},r_{3456} - r_{123456} ,
\nonumber \\
&&~~~~~~~
      r_4-r_{23456},r_{3456}-r_{23456},r_6-r_{23456},r_3-r_{3456},s_{36},s_{46} )
\nonumber \\
&& \nonumber \\
 &&+\kappa_{6345}
I^{(d)}_6(r_{123456},r_{23456},r_3,r_4,r_5,r_{3456};
      r_{23456}- r_{123456},r_3-r_{23456},s_{34},
 \nonumber \\
&&~~~~~~~
      s_{45},
      r_5-r_{3456},
      r_{3456} -r_{123456},
      r_3-r_{123456},r_4-r_{123456},r_5 - r_{123456} ,
      \nonumber \\
&&~~~~~~~    
      r_4-r_{23456},r_5-r_{23456},r_{3456}-r_{23456}, s_{35},
      r_3-r_{3456},r_4-r_{3456} ),
\nonumber \\
\label{I6step3}
\end{eqnarray}
where
\begin{eqnarray}
&&r_{3456}=-\frac{\lambda_{3456}}{g_{3456}},
~~~\kappa_{3456}=\frac{\partial r_{3456}}{\partial m_3^2},
~~~\kappa_{4356}=\frac{\partial r_{3456}}{\partial m_4^2},
\nonumber \\
&&
\kappa_{5346}=\frac{\partial r_{3456}}{\partial m_5^2},
~~~\kappa_{6345}=\frac{\partial r_{3456}}{\partial m_6^2}.
\end{eqnarray}
Functional relations for reducing other integrals  from the 
right-hand side of Eq.~(\ref{I6step2}) can be obtained
from Eq.~(\ref{I6step3}) by appropriate change of variables.

\begin{center}
{\it Reduction of the integral $I_6^{(d)}$,  step 4}
\end{center}

In the next step we derive formula for expressing integrals depending
on 9 variables in terms of integrals depending on 7 variables.
Again, as it was done in the previous step, we
substitute arguments of the first integral on the right-hand side 
of Eq.~(\ref{I6step3}) into Eq.~(\ref{I6_ini_fe}),  solve appropriate
systems of equations for the new unknowns and get
\begin{eqnarray}
&&I^{(d)}_6(r_{123456},r_{23456}, r_{3456},r_4,r_5,r_6;
      r_{23456}- r_{123456},r_{3456}-r_{23456},r_4-r_{3456},      
      \nonumber \\
&&~~~~~~~s_{45},s_{56},r_6 -r_{123456},    
      r_{3456}-r_{123456},r_4-r_{123456},r_5 - r_{123456} ,
      r_4-r_{23456},
\nonumber \\
&&~~~~~~~~r_5-r_{23456},
      r_6-r_{23456}, 
      r_5-r_{3456},r_6-r_{3456},s_{46} )
    \nonumber \\
&& \nonumber \\    
&&~=\kappa_{456}    
  I^{(d)}_6(r_{123456},r_{23456}, r_{3456},r_{456},r_5,r_6;
      r_{23456}- r_{123456},r_{3456}-r_{23456},r_{456}-r_{3456},      
      \nonumber \\
&&~~~~~~~r_5-r_{456},s_{56},r_6 -r_{123456},    
      r_{3456}-r_{123456},r_{456}-r_{123456},r_5 - r_{123456} ,
      r_{456}-r_{23456},
\nonumber \\
&&~~~~~~~~r_5-r_{23456},
      r_6-r_{23456}, 
      r_5-r_{3456},r_6-r_{3456}, r_6-r_{456})
    \nonumber \\
&& \nonumber \\    
    &&
+\kappa_{546}
I^{(d)}_6(r_{123456},r_{23456}, r_{3456},r_4,r_{456},r_6;
      r_{23456}- r_{123456},r_{3456}-r_{23456},r_4-r_{3456},      
      \nonumber \\
&&~~~~~~~
      r_4-r_{456},r_6- r_{456},r_6 -r_{123456},    
      r_{3456}-r_{123456},r_4-r_{123456},
      r_4-r_{23456},
\nonumber \\
&&~~~~~~~r_{456} - r_{123456},r_{456}-r_{23456},
      r_6-r_{23456}, 
      r_{456}-r_{3456},r_6-r_{3456},s_{46} )
    \nonumber \\
&& \nonumber \\    
&&+\kappa_{645}    
I^{(d)}_6(r_{123456},r_{23456}, r_{3456},r_4,r_5,r_{456};
      r_{23456}- r_{123456},r_{3456}-r_{23456},r_4-r_{3456},      
      \nonumber \\
&&~~~~~~~s_{45},r_5-r_{456},r_{456} -r_{123456},    
      r_{3456}-r_{123456},r_4-r_{123456},r_5 - r_{123456} ,
      r_4-r_{23456},
\nonumber \\
&&~~~~~~~~r_5-r_{23456},
      r_{456}-r_{23456}, 
      r_5-r_{3456},r_{456}-r_{3456},r_4-r_{456} ),
    \nonumber \\
\label{I6step4}    
\end{eqnarray}    
where
\begin{equation}
r_{456}=-\frac{\lambda_{456}}{g_{456}},
~~~\kappa_{456}=\frac{\partial r_{456}}{\partial m_4^2},
~~~\kappa_{546}=\frac{\partial r_{456}}{\partial m_5^2},
~~~\kappa_{645}=\frac{\partial r_{456}}{\partial m_6^2}.
\end{equation}
Notice that all integrals in the right-hand side of equation (\ref{I6step4}) 
depend on 7 variables.
These integrals may be expressed in terms of  integrals depending
on 6 variables.

\begin{center}
{\it Reduction of the integral  $I_6^{(d)}$,  step 5}
\end{center}

The final formula for the first integral on the right side
of Eq.~(\ref{I6step4})  was derived by the same method which was 
used in the previous steps and reads
\begin{eqnarray}
&&I^{(d)}_6(r_{123456},r_{23456}, r_{3456},r_{456},r_5,r_6;
      r_{23456}- r_{123456},r_{3456}-r_{23456},r_{456}-r_{3456},      
      \nonumber \\
&&~~~~~~~r_5-r_{456},s_{56},r_6 -r_{123456},    
      r_{3456}-r_{123456},r_{456}-r_{123456},r_5 - r_{123456} ,
      r_{456}-r_{23456},
\nonumber \\
&&~~~~~~~~r_5-r_{23456},
      r_6-r_{23456}, 
      r_5-r_{3456},r_6-r_{3456}, r_6-r_{456})
    \nonumber \\
&& \nonumber \\    
&&
=\kappa_{56}
I^{(d)}_6(r_{123456},r_{23456}, r_{3456},r_{456},r_{56},r_6;
      r_{23456}- r_{123456},r_{3456}-r_{23456},r_{456}-r_{3456},      
      \nonumber \\
&&~~~~~~~r_{56}-r_{456},r_6-r_{56},r_6 -r_{123456},    
      r_{3456}-r_{123456},r_{456}-r_{123456},r_{56} - r_{123456},     
\nonumber \\
&&~~~~r_{456}-r_{23456},r_{56}-r_{23456},
      r_6-r_{23456}, 
      r_{56}-r_{3456},r_6-r_{3456}, r_6-r_{456})
    \nonumber \\
&& \nonumber \\    
    &&
    +\kappa_{65}
    I^{(d)}_6(r_{123456},r_{23456}, r_{3456},r_{456},r_5,r_{56};
      r_{23456}- r_{123456},r_{3456}-r_{23456},r_{456}-r_{3456},      
      \nonumber \\
&&~~~~~~~r_5-r_{456},r_6-r_{56},r_{56} -r_{123456},    
      r_{3456}-r_{123456},r_{456}-r_{123456},r_5 - r_{123456} ,
\nonumber \\
&&~~~~r_{456}-r_{23456},r_5-r_{23456},
      r_{56}-r_{23456}, 
      r_5-r_{3456},r_{56}-r_{3456}, r_{56}-r_{456}),
\nonumber \\
\label{I6step5}
\end{eqnarray}
where
\begin{equation}
r_{56}=-\frac{\lambda_{56}}{g_{56}},
~~~~\kappa_{56}=\frac{\partial r_{56}}{\partial m_5^2},
~~~~\kappa_{65}=\frac{\partial r_{56}}{\partial m_6^2}.
\end{equation}
Analogous formulae for the reduction of other
integrals on the right-hand side of Eq.~(\ref{I6step4})
can be obtained form Eq.~(\ref{I6step5}) by  changing
variables appropriately. This completes the derivation 
of the reduction formulae for the integral $I_6^{(d)}$.

Composition of equations (\ref{I6step1}), (\ref{I6step2}),
(\ref{I6step3}), (\ref{I6step4}), (\ref{I6step5})  and all required relations 
that follow from these equations by changing variables
as mentioned  previously, gives a formula for the 
complete functional reduction of the integral $I_6^{(d)}$.
This formula represent integral depending on 21 variables
as a sum of 720 integrals, each of which depends only on 6
variables.
All resulting  integrals in this sum 
have the form
\begin{eqnarray}
&&
I_6^{(d)}(m^2_i,m^2_j,m^2_k,m^2_l,m^2_r,m^2_s;
~~m^2_j-m^2_i,m^2_k-m^2_j,m^2_l-m^2_k,m^2_r-m^2_l,
\nonumber \\
&&~~~~~
m^2_s-m^2_r,
m^2_s-m^2_i,
m^2_k-m^2_i,m^2_l-m^2_i,m^2_r-m^2_i,m^2_l-m^2_j,
\nonumber \\
&&~~~~~~~
m^2_r-m^2_j,m^2_s-m^2_j,
m^2_r-m^2_k,m^2_s-m^2_k,m^2_s-m^2_l),
\end{eqnarray}     
where $m^2_i$, $m^2_j$, $m^2_k$, $m^2_l$, $m^2_r$, $m^2_s$
are  ratios of polynomials in  masses and  kinematic invariants.
Coefficients in front of these integrals are also
ratios of polynomials in masses  and  kinematic invariants.

The final formula of the reduction is too lengthy to display here,
but we provide it
in a computer-readable ancillary file attached to this article.

\subsection{Dimensional
recurrence relation and series representation}

An analytic result for the integral $I_6^{(d)}$ can be obtained, for 
example,  by solving a dimensional recurrence relation or by evaluating
Feynman parameter integral.

Dimensional recurrence relation for the integral $I_6^{(d)}$
depending on the MNV reads
\begin{eqnarray}
&&(d-5)I^{(d+2)}_6(r_{123456},r_{23456}, r_{3456},r_{456},r_{56},r_6;
      r_{23456}- r_{123456},r_{3456}-r_{23456},r_{456}-r_{3456},      
      \nonumber \\
&&~~~~~~~r_{56}-r_{456},r_6-r_{56},r_6 -r_{123456},    
      r_{3456}-r_{123456},r_{456}-r_{123456},r_{56} - r_{123456},     
\nonumber \\
&&~~~~r_{456}-r_{23456},r_{56}-r_{23456},
      r_6-r_{23456}, 
      r_{56}-r_{3456},r_6-r_{3456}, r_6-r_{456})=
    \nonumber \\
&& \nonumber \\    
&&-2r_{123456}I^{(d)}_6(r_{123456},r_{23456}, r_{3456},r_{456},r_{56},r_6;
      r_{23456}- r_{123456},r_{3456}-r_{23456},r_{456}-r_{3456},      
      \nonumber \\
&&~~~~~~~r_{56}-r_{456},r_6-r_{56},r_6 -r_{123456},    
      r_{3456}-r_{123456},r_{456}-r_{123456},r_{56} - r_{123456},     
\nonumber \\
&&~~~~r_{456}-r_{23456},r_{56}-r_{23456},
      r_6-r_{23456}, 
      r_{56}-r_{3456},r_6-r_{3456}, r_6-r_{456})
    \nonumber \\
&& \nonumber \\    
&& -I^{(d)}_5(r_{23456}, r_{3456},r_{456},r_{56},r_6;
      r_{3456}- r_{23456},r_{456}-r_{3456},r_{56}-r_{456},      
      r_{6}-r_{56},r_6 -r_{23456},
      \nonumber \\
&&~~~~~~~
      r_{456}-r_{23456},r_{56}-r_{23456},
      r_{56}-r_{3456},
      r_{6} - r_{3456},r_6-r_{456}).
\end{eqnarray} 
Notice that the inhomogeneous term in this equation is
an integral depending on the MNV. 
Solution of this recurrence relation is straightforward but
cumbersome. The result is a bit lengthy and for this reason 
it will not be presented in this article.

We obtained the following Feynman parameter representation of
this integral 
\begin{eqnarray}
&&I_6^{(d)}(r_{123456},r_{23456},r_{3456},r_{456},r_{56},r_6;
\{s_{ij}\})
\nonumber \\
&&~~~~~~~~~~~~~~~~~~~~~~~~ =
\Gamma\left(6-\frac{d}{2}\right)
\int_0^1 ...\int_0^1~ x_1^4x_2^3x_3^2x_4~h_6^{\frac{d}{2}-6}  dx_1 ...dx_5,
\label{I6paramrep}
\end{eqnarray}
where
\begin{eqnarray}
&&h_6=
r_{123456}-(r_{123456}-r_{23456})x_1^2-(r_{23456}-r_{3456})x_1^2 x_2^2
-(r_{3456}-r_{456})x_1^2 x_2^2 x_3^2
\nonumber \\
&&~~~~~~~~~~~~~~~~
-(r_{456}-r_{56})x_1^2 x_2^2 x_3^2 x_4^2
-(r_{56}-r_6) x_1^2 x_2^2 x_3^2 x_4^2 x_5^2.
\label{h6genkin}
\end{eqnarray}
and
\begin{eqnarray}
&& s_{ij}=m_j^2-m_i^2, (j>i),
\nonumber \\
&&m_1^2=r_{123456},~m_2^2=r_{23456},~m_3^2=r_{3456},~m_4^2=r_{456},~
m_5^2=r_{56},~m_6^2=r_6.~~
\end{eqnarray}

Series representation of the integral $I_6^{(d)}$ can be obtained
by the same method which was used in deriving the 
series representation of integrals $I_3^{(d)}$,  $I_4^{(d)}$, $I_5^{(d)}$.
Expanding the integrand of (\ref{I6paramrep})  in terms
of variables
\begin{eqnarray}
&&z_1=\frac{r_{123456}-r_{23456}}{r_{123456}},~~
z_2=\frac{r_{123456}-r_{23456}}{r_{123456}},~~
z_3=\frac{r_{123456}-r_{23456}}{r_{123456}},~~
\nonumber \\
&&z_4=\frac{r_{123456}-r_{23456}}{r_{123456}},~~
z_5=\frac{r_{123456}-r_{23456}}{r_{123456}},
\end{eqnarray}
assuming that all $|z_j|<1$ and integrating 
over $x_1$,...,$x_5$ term by term, we get 
\begin{eqnarray}
&&I_6^{(d)}(m^2_{1},m^2_{2},m^2_{3},m^2_{4},m^2_{5},m^2_6;
\{s_{ij}\})
\nonumber \\
&&~~~~~=
\frac{r_{123456}^{\frac{d}{2}-6}}{120}
\Gamma\left(6-\frac{d}{2}\right) 
\sum_{n_1,n_2,n_3,n_4,n_5=0}^{\infty}\left(6-\frac{d}{2} 
\right)_{n_1+n_2+n_3+n_4+n_5}
\frac{(\frac52)_{n_1+n_2+n_3+n_4+n_5}}{(\frac72)_{n_1+n_2+n_3+n_4+n_5}}
\nonumber \\
&&~~~ \times
\frac{(2)_{n_2+n_3+n_4+n_5}}{(3)_{n_2+n_3+n_4+n_5}}
\frac{\left(\frac32\right)_{n_3+n_4+n_5}}{\left(\frac52\right)_{n_3+n_4+n_5}}
\frac{(1)_{n_4+n_5}}{(2)_{n_4+n_5}}
\frac{\left(\frac12 \right)_{n_5}}{\left(\frac32\right)_{n_5}}
\frac{z_1^{n_1}}{n_1!}\frac{z_2^{n_2}}{n_2!}\frac{z_3^{n_3}}{n_3!}
\frac{z_4^{n_4}}{n_4!}\frac{z_5^{n_5}}{n_5!}.
\end{eqnarray}

One can see that the summand of this series
is very similar to that of the series 
given in Eqs.~(\ref{I3_via_T2}), (\ref{T3series}), (\ref{I5viaT4}).

\section{Functional reduction of integrals with special kinematics}

The  procedure of functional reduction must be  modified if 
the Gram determinant $g_{12...n}=0$.  In this case the integral
$I_n^{(d)}$ can be reduced  \cite{Fleischer:1999hq} to a combination 
of integrals $I_{n-1}^{(d)}$. Functional reduction
can be applied to integrals obtained after such a reduction.
Notice that if the lower order Gram determinant $g_{ij..k}$ vanish 
then it means that the Gram determinant $g_{12...n}$ also vanish
\cite{byckling1973particle}.  
A modification of the functional reduction is needed
in case when some kinematic invariants $s_{ij}$ vanish.

If some $s_{ij}=0$ ($r_{ij}\rightarrow \infty$), then the corresponding 
last step of the functional reduction must be skipped.
There is no further functional reduction of integrals
with such values of kinematical invariants.
Analytic results for these integrals  are simpler
than those  for integrals depending on a general kinematics. 
We will consider derivation of these results integral by integral.

\begin{center}
{\it The integral $I_3^{(d)}$  at $s_{23}=0$}
\end{center}

If, at the last step of functional reduction,
the kinematic invariant of the integral $I_3^{(d)}$, 
 say  $s_{23}=0$, then the application of the  
formula (\ref{I3step2}) should be skipped.

The Feynman parameter  representation of this integral reads
\begin{eqnarray}
&&
I_3(r_{123},r_2,r_3;~0,r_3-r_{123},r_2-r_{123})
\nonumber\\
&& = 
-\Gamma\left(3-\frac{d}{2}\right)
\int_0^1\int_0^1x_1 
[r_{123}-(r_{123}-r_3)x_1^2-(r_3-r_2)x_1^2x_2]^{\frac{d}{2}-3}dx_1dx_2.~~~~
\label{I3_spec_case}
\end{eqnarray}
Notice a small difference between this expression and the
Feynman parameter representation of the integral $I_3^{(d)}$
given in Eq.~(\ref{I3param}).
The integral (\ref{I3_spec_case}) can easily be evaluated.
First, integrating with respect to $x_1$ and then 
with respect to $x_2$ yields
\begin{eqnarray}
&&I_3^{(d)}(r_{123},r_{2},r_3;~0,
r_3-r_{123},r_{2}-r_{123})=
\frac{- \pi}{2\sin \frac{\pi d}{2} \Gamma\left(\frac{d}{2}-1\right) }
\nonumber \\
&&
\times 
\left\{ \frac{r_3^{\frac{d}{2}-2}}
{r_3-r_{123}}
F_1\left(1,1,2-\frac{d}{2},2; \frac{r_3-r_2}{r_3-r_{123}},\frac{r_3-r_2}{r_3}\right)
-\frac{r_{123}^{\frac{d}{2}-2}}{r_2-r_3} 
\ln\frac{r_{123}-r_2}{r_{123}-r_3}
\right\}.~~~
\label{i3_zero_int}
\end{eqnarray}
The result for this integral may be obtained in a slightly different form.
Expanding the integrand of (\ref{I3_spec_case}) with respect to variables
\begin{equation}
z_1=\frac{r_{123}-r_3}{r_{123}},~~~~~z_2=\frac{r_3-r_2}{r_{123}},
\end{equation}
assuming $|z_1|<1$, $|z_2|<1$ and integrating with respect to  $x_1$ and $x_2$
term by term, we get the series representation
\begin{eqnarray}
&&I_3^{(d)}(r_{123},r_{2},r_3;~0,r_3-r_{123},r_{2}-r_{123})
\nonumber \\
&&
=-\frac{r_{123}^{\frac{d}{2}-3}}{2}\Gamma\left(3-\frac{d}{2}\right)
\sum_{n_1,n_2=0}^{\infty}\left(3-\frac{d}{2}\right)_{n_1+n_2}
\frac{(1)_{n_1+n_2}}{(2)_{n_1+n_2}}
\frac{(1)_{n_2}}{(2)_{n_2}} \frac{z_1^{n_1}}{n_1!}\frac{z_2^{n_2}}{n_2!}.
\label{I3series_spec}
\end{eqnarray}

Another series representation can be obtained by a slight 
modification of the above derivation. We  expand 
the integrand  with respect to   $z_1$ and integrate over $x_1$
first.
Finally, performing the  integration with respect to $x_2$, we get result
in terms of two generalized hypergeometric series $_3F_2$
\begin{eqnarray} 
&&I_3^{(d)}(r_{123},r_{2},r_3;~0,r_3-r_{123},r_{2}-r_{123})
=-\frac{r_{123}^{\frac{d}{2}-3}}{2}\Gamma\left(3-\frac{d}{2}\right)
\nonumber \\
&&~~ \times
\left\{\frac{z_1+z_2}{z_2} \Fh32\FeZ{3-\frac{d}{2},1,1}{2,2}
- \frac{z_1}{z_2} \Fh32\FEZ{3-\frac{d}{2},1,1}{2,2}
\right \}.
\label{I3_via_f32}
\end{eqnarray}

Analytical expression for the integral can also be obtained  by 
solving dimensional recurrence relation
\begin{eqnarray}
&&(d-2)I_3^{(d+2)}(r_{123},r_{2},r_3;~0,r_3-r_{123},r_{2}-r_{123})
=
\nonumber \\
&&~~~~~~~~~~~~-2 r_{123}I_3^{(d)}(r_{123},r_{2},r_3;~0,
r_3-r_{123},r_{2}-r_{123})
-I_2^{(d)}(r_{2},r_3;0).~~~~~~~
\label{dim_rec_i3zero}
\end{eqnarray}
Here the integral $I_2^{(d)}$ is a combination of two tadpole
integrals. Solution of this dimensional recurrence relation reads
\begin{eqnarray}
&&I_3^{(d)}(r_{123},r_{2},r_3;~0,
r_3-r_{123},r_{2}-r_{123})
=
\nonumber \\
&&
\frac{\pi r_{123}^{\frac{d}{2}-2}}{(r_2-r_3)\sin \frac{\pi d}{2} 
\Gamma \left(\frac{d}{2}\right)}
\left\{
\frac{d-2}{4}\ln(r_{123}-r_2)
-\frac{d-2}{4}\ln(r_{123}-r_3)
\right.
\nonumber \\
&&\left. +\frac12 \left(\frac{r_2}{r_{123}}\right)^{\frac{d}{2}-1}
\Fh21\Frd{1,\frac{d-2}{2}}{\frac{d}{2}}
- \frac12 \left(\frac{r_3}{r_{123}}\right)^{\frac{d}{2}-1}
\Fh21\Frt{1,\frac{d-2}{2}}{\frac{d}{2}}
\right\}.
\label{solu_i3_zero}
\end{eqnarray}
An arbitrary function, invariant under  $d\rightarrow d+2$, 
appearing in the  solution of  Eq.~(\ref{dim_rec_i3zero}), 
was obtained by solving a system of differential equations
with respect to kinematic variables.

As was shown in ref.~\cite{Kniehl:2011ym}, different representations 
of Feynman integrals can be used to find  new relations between 
hypergeometric functions. In particular, a comparison 
of Eq.~(\ref{I3_via_f32}) with Eq.~(\ref{solu_i3_zero}) yields
the following relationship
\begin{eqnarray}
&&\Fh32\Fz{3-\frac{d}{2},1,1}{2,2}=
\nonumber \\
&&~~~~~~~~~
\frac{2}{z(d-4)}
\left\{\frac{2(1-z)^{\frac{d-2}{2}}}{d-2}\Fh21\Fzm{1,\frac{d-2}{2}}{\frac{d}{2}}
+\ln z+\psi\left(\frac{d}{2}-1\right)+\gamma
\right\},~~~~~
\end{eqnarray}
where function $ \psi(x)$ is the logarithmic derivative of the Euler's 
$\Gamma$ function, $\psi(x)=d~ \ln \Gamma(x)/dx$, and $\gamma=0.57721566...$ denotes
Euler's or Mascheroni's constant \cite{Erdelyi:1953:HTF1}.

By comparing Eq.~(\ref{i3_zero_int}) with  Eq.~(\ref{solu_i3_zero}), we get 
the following relationship
\begin{eqnarray}
&&F_1\left(1,1,2-\frac{d}{2},2;x,y\right)=
\nonumber \\
&&~~~~~~~~~~~~
\frac{2}{(d-2)y}
 \Biggl\{\!\frac{(1-y)^{\frac{d}{2}-1}}{x-1}
 \Fh21\FRD{1,1}{\frac{d}{2}}
\!+\!\Fh21\FRT{1,1}{\frac{d}{2}}\!
\Biggr\}.~~
\label{F1_via_F21}
\end{eqnarray}
This formula can be used to evaluate a high-order
series expansion in $\varepsilon =(4-d)/2$ of the hypergeometric
Appell function $F_1$. Such an expansion can easily be derived,
as expansion of the hypergeometric functions $_2F_1$ 
from Eq.~(\ref{F1_via_F21}) is known to any order
in $\varepsilon$ .

\begin{center}
{\it The integral $I_4^{(d)}$  at $s_{34}=0$}
\end{center}

The situation concerning the integral $I_4^{(d)}$ is very similar to the 
case of the integral $I_3^{(d)}$.
If, for example, a kinematic variable of the integral $I_4^{(d)}$
in equation (\ref{I4step3}), say $s_{34}=0$,
then the application of the relation (\ref{I4step3}) must be skipped.

Analytic result for the integral  $I_4^{(d)}$  in this case 
can  also be obtained by solving a dimensional recurrence relation
\begin{eqnarray}
&&
(d-3)   I_4^{(d+2)}(r_{1234},r_{234},r_{34},r_4; 
\nonumber \\
&&~~~~~~~
   r_{234}-r_{1234},r_{34}-r_{234},0,r_4-r_{1234},r_4-r_{234},r_{34}-r_{1234})
\nonumber \\  
&& \nonumber \\
&&~~  =  -2r_{1234} I_4^{(d)}(r_{1234},r_{234},r_{34},r_4; 
\nonumber \\
&&~~~~~~~
   r_{234}-r_{1234},r_{34}-r_{234},0,r_4-r_{1234},r_4-r_{234},r_{34}-r_{1234})
\nonumber \\  
&& \nonumber \\
&&~~~~~~~~~~~~~~
- I_3^{(d)}(r_{234}, r_{34}, r_4; 0, r_4 - r_{234}, r_{34} - r_{234} ).
\end{eqnarray}
In order to solve this equation we used the analytic result
(\ref{solu_i3_zero}) for  the  integral $I_3^{(d)}$ and obtained
\begin{eqnarray}
&&   I_4^{(d)}(r_{1234},r_{234},r_{34},r_4; 
\nonumber \\
&&~~~~~~~
   r_{234}-r_{1234},r_{34}-r_{234},0,r_4-r_{1234},r_4-r_{234},r_{34}-r_{1234})
\nonumber \\  
&&=
\frac{ r_{1234}^{\frac{d}{2}} ~c_4(r_{1234},r_{234},r_{34},r_4)}
{ \Gamma\left(\frac{d-3}{2}\right) \sin{\frac{\pi d}{2}}}
-\frac{\pi}{8\sin{\frac{\pi d}{2}}\Gamma\left(\frac{d}{2}\right)~r_{1234}(r_{34}-r_4)}
\nonumber \\
&&\times \Biggl\{(d-2) r_{234}^{\frac{d}{2}-2}~
\Bigl[\ln(r_{234}-r_{34})-\ln(r_{234}-r_4)\Bigr]
          \Fh21\FOO{1,\frac{d-3}{2}}{\frac{d-2}{2}}
\nonumber \\     
&&
 +\frac{2 r_{34}^{\frac{d}{2}-1}}{r_{234}-r_{34}}
 F_3\left(1,1,1,\frac{d-3}{2},\frac{d}{2}; 
 \frac{r_{34}}{r_{34}-r_{234}},\frac{r_{34}}{r_{1234}} \right)
\nonumber
\\
&&
- \frac{2r_4^{\frac{d}{2}-1}}{r_{234}-r_4}
 F_3\left(1,1,1,\frac{d-3}{2},\frac{d}{2}; 
 \frac{r_{4}}{r_{4}-r_{234}},\frac{r_4}{r_{1234}}     \right) \Biggr\},
\end{eqnarray} 
where  
\begin{eqnarray}
&&c_4(r_{1234},r_{234},r_{34},r_4)=
\nonumber \\
&&~~
\frac{\pi^{\frac32}}{4 r_{1234}^{\frac52}(r_{34}-r_4)
(r_{1234}-r_{234})^{\frac12}  }
\Biggl\{\ln(r_{234}-r_{34}) -\ln(r_{234}-r_4)
\nonumber 
\\
&&
-\ln\frac{\sqrt{r_{1234}-r_{34}}-\sqrt{r_{1234}-r_{234}}}
{\sqrt{r_{1234}-r_{34}}+\sqrt{r_{1234}-r_{234}}}
+\ln\frac{\sqrt{r_{1234}-r_{4}}-\sqrt{r_{1234}-r_{234}}}
{\sqrt{r_{1234}-r_{4}}+\sqrt{r_{1234}-r_{234}}}
\Biggr\}.
\end{eqnarray}
The definition of the  hypergeometric Appell function  $F_3$ is given 
in Appendix.
The function $c_4(r_{1234},r_{234},r_{34},r_4)$ appeared as an
arbitrary periodic function in the solution of the dimensional recurrence relation 
for the integral $I_4^{(d)}$. This function was obtained by solving the 
system of differential equations which was derived from a system of 
differential equations for the integral $I_4^{(d)}$.

The integral $I_4^{(d)}$ can also be evaluated using the 
Feynman parameter representation.
In the case under consideration, the integral representation 
is slightly different from the representation given in (\ref{I4_param_rep}) 
\begin{eqnarray}
&&I_4^{(d)}(r_{1234},r_{234},r_3,r_4; r_{234}-r_{1234},r_3-r_{234},0,
r_4-r_{1234},r_4-r_{234},r_3-r_{1234})
\nonumber \\
&& \nonumber \\
&&~~~~~~~~~~~~~~~~~~~~~~~~
=\Gamma\left(4-\frac{d}{2}\right)
\int_0^1\int_0^1\int_0^1 x_1^2x_2h_4^{\frac{d}{2}-4}dx_1dx_2dx_3,
\label{I4z_param}
\end{eqnarray}
where
\begin{equation}
h_4=r_{1234}-(r_{1234}-r_{234})x_1^2 -(r_{234}-r_3)x_1^2x_2^2
-(r_3-r_4)x_1^2x_2^2x_3.
\end{equation}
The difference between this $h_4$ here and the $h_4$ from Eq.~(\ref{h4})
is only in the last term.
Expanding the integrand of (\ref{I4z_param})
with respect to the three variables
\begin{equation}
z_1=\frac{r_{1234}-r_{234}}{r_{1234}},~~~
z_2=\frac{r_{234}-r_{3}}{r_{1234}},~~~
z_3=\frac{r_{3}-r_{4}}{r_{1234}},
\end{equation}
assuming that $|z_j|<1,~~(j=1,2,3)$ 
integrating with respect to $x_1$, $x_2$, $x_3$ term by term, we then obtain 
\begin{eqnarray}
&&   I_4^{(d)}(r_{1234},r_{234},r_{34},r_4; 
 r_{234}-r_{1234},r_{34}-r_{234},0,
\nonumber \\
&&~~~~~~~
  r_4-r_{1234},r_4-r_{234},r_{34}-r_{1234})=
\frac16 \Gamma\left(4-\frac{d}{2}\right)r_{1234}^{\frac{d}{2}-4}
\nonumber \\
&&~~\times 
\sum_{n_1,n_2,n_3=0}^{\infty}
\left(4-\frac{d}{2}\right)_{n_1+n_2+n_3}
\frac{\left(\frac32\right)_{n_1+n_2+n_3}}
{\left(\frac52\right)_{n_1+n_2+n_3}}
\frac{\left( 1 \right)_{n_2+n_3}}{\left( 2\right)_{n_2+n_3}}
\frac{\left(1\right)_{n_3}}{\left(2\right)_{n_3}}
\frac{z_1^{n_1} z_2^{n_2}z_3^{n_3}}{n_1!n_2!n_3!}.
\label{I4series_s34_zero}
\end{eqnarray}
Note that the summand of this series is slightly different
from the summand  of the integral $I_4^{(d)}$ for a
general kinematics in Eq.~(\ref{T3series}).

\begin{center}
{\it The integral $I_5^{(d)}$  at $s_{45}=0$}
\end{center}

If one of the kinematic variables of the integral $I_5^{(d)}$, say
$s_{45}=0$, then the application of the reduction relation
(\ref{I5step4}) must be skipped.

The analytic result for such an integral can be obtained either
by solving a dimensional recurrence relation or by calculating
Feynman parameter integral.
Solving the dimensional recurrence relation for the integral
$I_5^{(d)}$ is somewhat cumbersome and the result is relatively long.
For these reasons, we will not present it here. 
Instead, we have
derived expression for the integral $I_5^{(d)}$ in terms of multiple 
hypergeometric series. The Feynman parameter representation
of this integral reads
\begin{eqnarray}
&&I_5^{(d)}(r_{12345},r_{2345},r_{345},r_4,r_5;  r_{2345}-r_{12345},r_{345}-r_{2345},
r_4-r_{345},0,r_5-r_{12345},
\nonumber \\
&&~~~~~~~~~~~~~~~~~~~~
r_{345}-r_{12345},r_4-r_{12345},r_4-r_{2345} , r_5-r_{2345},r_5-r_{345})
\nonumber \\
&&~~~~~~~~~ = -\Gamma\left(5-\frac{d}{2}\right)
\int_0^1\int_0^1\int_0^1 \int_0^1 x_1^3x_2^2x_3h_5^{\frac{d}{2}-5}dx_1dx_2dx_3dx_4,
\end{eqnarray}
where
\begin{eqnarray}
&&
h_5=r_{12345}-(r_{12345}- r_{2345})x_1^2
\nonumber \\
&&~~~~~~
-(r_{2345}-r_{345}) x_1^2x_2^2
-(r_{345}-r_4) x_1^2 x_2^2 x_3^2-(r_4-r_5) x_1^2 x_2^2 x_3^2 x_4.~~
\end{eqnarray}
Expanding the Feynman parameter integrand with respect to the four variables
\begin{equation}
z_1=\frac{r_{12345}- r_{2345}}{r_{12345}},~~~
z_2=\frac{r_{2345}- r_{345}}{r_{12345}},~~~
z_3=\frac{r_{345}- r_{4}}{r_{12345}},~~~
z_4=\frac{r_{4}- r_{5}}{r_{12345}},~~~
\end{equation}
and integrating over  $x_1$,...,$x_4$ term by term then yields 
\begin{eqnarray}
&&I_5^{(d)}(r_{12345},r_{2345},r_{345},r_4,r_5; r_{2345}-r_{12345},r_{345}-r_{2345},
r_4-r_{345},0,r_5-r_{12345},
\nonumber \\
&&~~~~~~
r_{345}-r_{12345},r_4-r_{12345},r_4-r_{2345} , r_5-r_{2345},r_5-r_{345})
\nonumber \\
&&~~~~ =
-\frac{1}{24}\Ga{5-\frac{d}{2}}
\sum_{n_1,n_2,n_3,n_4=0}^{\infty}\left(5-\frac{d}{2}\right)_{n_1+n_2+n_3+n_4}
\nonumber \\
&&~~~~~~~~~~~~~\times
\frac{(2)_{n_1+n_2+n_3+n_4}}{(3)_{n_1+n_2+n_3+n_4}}
\frac{\left(\frac32\right)_{n_2+n_3+n_4}}{\left(\frac52\right)_{n_2+n_3+n_4}}
\frac{(1)_{n_3+n_4}}{(2)_{n_3+n_4}}
\frac{\left(1 \right)_{n_4}}{\left(2\right)_{n_4}}
\frac{z_1^{n_1}}{n_1!}\frac{z_2^{n_2}}{n_2!}\frac{z_3^{n_3}}{n_3!}
\frac{z_4^{n_4}}{n_4!}.
\label{I5series_s45_zero}
\end{eqnarray}
Note a slight difference between this representation
and series representation (\ref{I5viaT4})
of the integral $I_5^{(d)}$ for a general kinematics.

\begin{center}
{\it The integral $I_6^{(d)}$  at $s_{56}=0$}
\end{center}

The integral $I_6^{(d)}$ for the case when  one of the 
kinematic variables, say $s_{56}=0$, should be 
considered in the same way as the integral $I_5^{(d)}$.
The application  of the reduction formula (\ref{I6step5})
in the last step must be skipped.

The analytic calculation of the integral can be performed either by solving
the dimensional recurrence relation or by evaluating the Feynman
parameter integral. The solution of the dimensional recurrence relation
is straightforward but cumbersome  and the result is relatively
long. For this reason the derivation of the solution and the 
result will not  be considered in the present paper.

The parametric representation of the integral reads
\begin{eqnarray}
&&I_6^{(d)}(r_{123456},r_{23456},r_{3456},r_{456},r_5,r_6;
       r_{23456}-r_{123456},r_{3456}-r_{23456},r_{456}-r_{3456},~~~~       
\nonumber \\
&&~~~~~~~~r_5-r_{456},0,  r_6-r_{123456} ,r_{3456}-r_{123456},
      r_5-r_{123456}, r_{456}-r_{23456},
  \nonumber \\
&&~~~~~~~~r_{456}-r_{123456}, r_5-r_{23456},
      r_6-r_{23456},r_5-r_{3456},
r_6-r_{3456},r_6-r_{456})
\nonumber \\
&& \nonumber \\
&&~~~~~~~~~~~~ = \Gamma\left(6-\frac{d}{2}\right)
\int_0^1 ... \int_0^1
x_1^4x_2^3x_3^2x_4 h_6^{\frac{d}{2}-6}  dx_1dx_2dx_3dx_4dx_5,
\end{eqnarray}
where
\begin{eqnarray}
&&h_6=
r_{123456}-(r_{123456}-r_{23456})x_1^2-(r_{23456}-r_{3456})x_1^2 x_2^2
\nonumber \\
&&~~~~
-(r_{3456}-r_{456})x_1^2 x_2^2 x_3^2
-(r_{456}-r_5)x_1^2 x_2^2 x_3^2 x_4^2-(r_5-r_6) x_1^2 x_2^2 x_3^2 x_4^2x_5.
\end{eqnarray}
The last term of $h_6$ here differs  from that given
in Eq.~(\ref{h6genkin}).
Expanding the integrand with respect to the five variables 
\begin{eqnarray}
&&
z_1=\frac{r_{123456}-r_{23456}}{r_{123456}},~
z_2=\frac{r_{23456}-r_{3456}}{r_{123456}},~
\nonumber \\
&&
z_3=\frac{r_{3456}-r_{456}}{r_{123456}},~
z_4=\frac{r_{456}-r_{5}}{r_{123456}},~
z_5=\frac{r_{5}-r_{6}}{r_{123456}},~
\end{eqnarray}
and integrating over $x_1$,...,$x_5$ term by term, we get the
multiple series representation 
\begin{eqnarray}
&&I_6^{(d)}(r_{123456},r_{23456},r_{3456},r_{456},r_5,r_6; 
       r_{23456}-r_{123456},r_{3456}-r_{23456},r_{456}-r_{3456},~~~~       
\nonumber \\
&& ~~~~~~r_5-r_{456},0,  r_6-r_{123456} ,r_{3456}-r_{123456},r_{456}-r_{123456},
      r_5-r_{123456}, r_{456}-r_{23456},
  \nonumber \\
  &&~~~~    
       r_5-r_{23456},
      r_6-r_{23456},r_5-r_{3456},
r_6-r_{3456},r_6-r_{456})
\nonumber \\
&& \nonumber \\
&&~~~=
\frac{r_{123456}^{\frac{d}{2}-6}}{120}
\Gamma\left(6-\frac{d}{2}\right) 
\sum_{n_1,n_2,n_3,n_4,n_5=0}^{\infty}\!\!\left(6-\frac{d}{2} 
\right)_{n_1+n_2+n_3+n_4+n_5}
\frac{(\frac52)_{n_1+n_2+n_3+n_4+n_5}}{(\frac72)_{n_1+n_2+n_3+n_4+n_5}}
\nonumber \\
&&~~~ \times
\frac{(2)_{n_2+n_3+n_4+n_5}}{(3)_{n_2+n_3+n_4+n_5}}
\frac{\left(\frac32\right)_{n_3+n_4+n_5}}{\left(\frac52\right)_{n_3+n_4+n_5}}
\frac{(1)_{n_4+n_5}}{(2)_{n_4+n_5}}
\frac{\left(1 \right)_{n_5}}{\left(2\right)_{n_5}}
\frac{z_1^{n_1}}{n_1!}\frac{z_2^{n_2}}{n_2!}\frac{z_3^{n_3}}{n_3!}
\frac{z_4^{n_4}}{n_4!}\frac{z_5^{n_5}}{n_5!}.
\end{eqnarray}
Note similarities between summand of this multiple
series and the summands of (\ref{I3series_spec}),
(\ref{I4series_s34_zero}), (\ref{I5series_s45_zero}).

This concludes our consideration of integrals for special values of 
kinematic variables.

\section{General algorithm of the functional reduction}
 
Comparing  expressions 
(\ref{I2funcred}), (\ref{I3funcred}), (\ref{I4_final}),
(\ref{I5step1}), (\ref{I5step2}), (\ref{I5step3}), (\ref{I5step4}),
(\ref{I6step1}), (\ref{I6step2}), (\ref{I6step3}), (\ref{I6step4}), (\ref{I6step5}),
it is not hard to see common features and similarities between them.
Based on these observations, 
we have developed a regular  algorithm for obtaining final 
reduction formulae, which is valid for the  integrals 
considered in the article.
We  assume that the algorithm can be applied  to  integrals 
$I_n^{(d)}$ with $n>6$ as well.

Final functional reduction formulae for the integrals 
$I_2^{(d)}$, {\ldots}, $I_6^{(d)}$
can be obtained by exploiting the following algorithm:
\begin{itemize}
\item
write down the term
\begin{equation}
\kappa_{1{\ldots} n} \kappa_{2 {\ldots} n}{\ldots} 
\kappa_{n-1~n}~~~
I_n^{(d)}(m^2_1, m^2_2,{\ldots} m_n^2; s_{12},s_{23},{\ldots} )
\label{initerm}
\end{equation}
\end{itemize}
\begin{itemize}
\item replace in the integral {
$s_{ij} \rightarrow m_j^2-m_i^2$ } $(j>i)$
\item replace in the integral {
$m_1^2 \rightarrow r_{1{\ldots} n}$,
 $m_2^2 \rightarrow r_{2{\ldots} n } , {\ldots}, m_n^2 \rightarrow r_n $}  
\item
replace 
{
$\kappa_{i j  {\ldots} } \rightarrow  \frac{\partial r_{i j
{\ldots}}}{\partial m^2_i} $}
\item
generate {
$n!-1$} terms by
symmetrizing  the term (\ref{initerm}) with respect to the indices 
$1,2,{\ldots} n$ and  add all these terms to (\ref{initerm}).
\end{itemize}
All steps are very straightforward and easily 
achieved with a computer program. 
This algorithm works perfectly for integrals 
$I_2^{(d)}$,...,$I_6^{(d)}$.
We  verified numerically  that it is also valid 
for integrals $I_7^{(d)}$,  $I_8^{(d)}$.
Notice that  the  number of terms in the final reduction formula 
for massless integrals is $n!/2$.

We found that the parametric representation of the integral $I_n^{(d)}$ 
depending on the MNV can be written as
\begin{eqnarray}
&&
I_n^{(d)}(m_1^2,...,m_n^2; \{s_{ik}=m_k^2-m_i^2 | k>i \}) 
\nonumber \\
&&~~=
\frac{(-1)^nr_{1...n}^{\frac{d}{2}-n}}
{2^{n-1}}\Gamma\left(n-\frac{d}{2}\right)
\int_0^1 \frac{dt_1}{\sqrt{t_1}}\int_0^{t_1}
\frac{dt_2}{\sqrt{t_2}}~...~\int_0^{t_{n-2}}
\frac{h_n^{\frac{d}{2}-n} }{\sqrt{t_{n-1}}}
~ dt_{n-1}, 
\label{In_via_Tn}
\end{eqnarray}
where $h_n$ is a polynomial linear in the integration variables
\begin{equation}
h_n=1-z_1t_1-z_2t_2-...-z_{n-1}t_{n-1},~~~~~~~~z_i=\frac{m_i^2-m_{i+1}^2}{m_1^2}.
\label{h_n}
\end{equation}

The parametric representation of  integrals depending on the 
special kinematics
considered in Section 9  differs from that of (\ref{In_via_Tn}),
and reads
\begin{eqnarray}
&&
I_n^{(d)}(m_1^2,...,m_n^2; \{ s_{n-1,n}=0; s_{ik}=m_k^2-m_i^2 |{i<k} \})
\nonumber \\
&&~~~~~
=
\frac{(-1)^nr_{1...n}^{\frac{d}{2}-n}}
{2^{n-2}}\Gamma\left(n-\frac{d}{2}\right)
\nonumber \\
&& ~~\times 
\int_0^1 \frac{dt_1}{\sqrt{t_1}}\int_0^{t_1}
\frac{dt_2}{\sqrt{t_2}}~...~\int_0^{t_{n-4}}  \frac{dt_{n-3}}{\sqrt{t_{n-3}}} 
 \int_0^{t_{n-3}}   \frac{dt_{n-2}}{t_{n-2}}   \int_0^{t_{n-2}}
h_n^{\frac{d}{2}-n}
~ dt_{n-1},
\label{In_via_Tn_spec}
\end{eqnarray}
where $h_n$ is given in (\ref{h_n}). The integration with
respect to $t_{n-1}$ can be performed explicitly.  As a result 
of this integration, the integral $I_n^{(d)}$ that depends on $n-1$
variables, will be
expressed as a difference of two functions, each of which
depends on  $ n-2 $ variables. In section 9 such a 
representation was derived for the integrals $I_3^{(d)}$,  $I_4^{(d)}$.

Multiple series representation of the integral $I_n^{(d)}$ for $n=2,...,6$
was given in the previous sections.
The generic form of all these series  is 
\begin{eqnarray}
&&T_k(a,\{\beta_j\};\{\gamma_i\}; z_1,...,z_k)=
\nonumber \\
&&~~~~~\sum_{n_1,...,n_k=0}^{\infty}\left( a \right)_{n_1+n_2...+n_k}
\frac{ (\beta_1)_{n_1+n_2...+n_k}}{(\gamma_1)_{n_1+n_2...+n_k}}
~
\frac{(\beta_2)_{n_2+...+n_k}}{(\gamma_2)_{n_2+...+n_k}}
...\frac{(\beta_k)_{n_k}}{(\gamma_k)_{n_k}}
\frac{z_1^{n_1}}{n_1!}...\frac{z_k^{n_k}}{n_k!}.~~~~~~
\label{T_k_function}
\end{eqnarray}
This representation holds for integrals depending on general
kinematics 
as well as for integrals depending on the special kinematics 
considered in section 9.
As one can see from the considered examples,
the integral $I_n^{(d)}$ depending on a general kinematics 
can be written in terms of the function $T_{n-1}$ with  parameters
\begin{equation}
a=n-\frac{d}{2},~~~\beta_k=\frac{n-k}{2},~~~\gamma_k=\beta_k+1,~~~ 1\leq k \leq n-1.
\label{bk_gk}
\end{equation}
The integral $I_n^{(d)}$ depending on the special  kinematics 
can be written in terms of the function $T_{n-1}$ with  parameters
\begin{eqnarray}
&&a=n-\frac{d}{2},~~~~\beta_{n-1}=1,~~~~\beta_k=\frac{n-k}{2},
~~~~1\leq k \leq n-2;
\nonumber \\
&&\gamma_j=\beta_j+1,~~~~1\leq j \leq n-1.
\label{bk_gk_spec}
\end{eqnarray}
We assume that for $n>6$  the integrals $I_n^{(d)}$ 
depending on a generic as well as special kinematics
can also  be expressed in terms of the hypergeometric series
given in (\ref{T_k_function}) with parameters $\beta_k$, $\gamma_k$
defined in (\ref{bk_gk}), (\ref{bk_gk_spec}).

Note that the functions  $T_1$ and $T_2$ can be identified with 
the already known hypergeometric functions $_2F_1$ and  $S_1$:
\begin{eqnarray}
&&T_1(a,\beta_1;\gamma_1;z_1)=\Fh21\FzZ{a,\beta_1}{\gamma_1},
\nonumber \\
&&T_2(a,\beta_1,\beta_2;\gamma_1,\gamma_2;z_1,z_2)=
S_1(a,\beta_1,\beta_2,\gamma_1,\gamma_2;z_2,z_1).
\end{eqnarray}
The function $T_k(a,\{\beta_i\}; \{\gamma_j\}; \{z_n\})$
can be considered as a generalization 
of hypergeometric functions $S_1$ and $_2F_1$.

At present, there are several publications
where series representations of one-loop integrals
were considered.
In ref.~\cite{Kershaw:1973km}, it has been
shown that the $n$-point one-loop integral can be represented by a 
generalized hypergeometric power series depending on 
$n(n-1)/2$ variables. 
In refs. \cite{Davydychev:1990jt}, \cite{Davydychev:1990cq}
a representation of a general
scalar $n$-point one-loop Feynman integral 
in terms of $n(n+1)/2$-fold multiple hypergeometric series was
derived by using Mellin-Barnes technique.

We expect that our representation of one-loop integrals in terms of $(n-1)$-fold 
hypergeometric series will be  useful  for the analytic 
continuations as well as for the  $\varepsilon$ expansion of one-loop
integrals.

\section{Conclusions and outlook}

In this paper, we provided a systematic approach for reducing a
generic $n$-point one-loop integral with arbitrary masses
and kinematic invariants to a linear combination of integrals 
that depend on $n$ variables. The integrals  depending on the MNV
encountered at the last stage of the reduction were expressed in terms
of multiple hypergeometric series depending on $n-1$ dimensionless
variables. We have not found functional relations allowing for a
further reduction in the number of variables.

We have shown that analytic results for integrals with the MNV 
can be derived  by solving dimensional recurrence relations.
Explicit expressions for the integrals $I_2^{(d)}$,
$I_3^{(d)}$, $I_4^{(d)}$ as  solutions of dimensional recurrence
relations were given. Arbitrary periodic 
functions appearing in the solutions of dimensional recurrence relations
were found by solving systems of differential equations.

The choice of integrals depending on the MNV is not unique.
One can find relationships between integrals depending
on different minimal sets of variables using our functional relations
and rewrite results in a most preferable set of functions.
In section 4 such relationship was given  for the integral $I_2^{(d)}$.
Relevant relationship for the integral $I_3^{(d)}$ was presented 
in ref.~\cite{Kniehl:2011ym} and analogous relationships will be 
given for other integrals in a forthcoming publication.

We expect that our representation of one-loop integrals
can be helpful for deriving $\varepsilon=(4-d)/2$ expansion
of these integrals. For instance, 
multiple  series  (\ref{T_k_function}) 
can be expanded in $\varepsilon$ by exploiting the methods proposed
in refs.~\cite{bytev:2020zhg},\cite{Blumlein:2021hbq} or by solving 
system of differential equations for this series. In the latter case, 
to effectively solve  the problem, one should construct an appropriate 
alphabet. As shown in ref.~\cite{Chen:2022fyw}
the alphabet for the one-loop integrals can be expressed in terms of
the Gram determinants. We expect that our representation
of integrals  in terms of multiple hypergeometric series
those arguments   depend 
explicitly on the Gram determinants, can be useful
for finding canonical basis used to solve a system of 
differential equations as well as for finding an
alphabet of these integrals.

The new set of hypergeometric series $T_k$, encountered in 
computation of integrals depending on the MNV, will be 
studied in detail in our future publications. 

We plan to formulate a systematic procedure based on functional
relations that would allow analytic continuation of Feynman 
integrals to different kinematic domains.
As it was discovered in the course of our preliminary 
investigation (see also \cite{Kniehl:2011ym}), the functional relations 
can help to find still unknown relationships between 
hypergeometric functions.

We also plan to apply the functional reduction method
for evaluating Feynman diagrams required for
computing radiative corrections for modern experiments.

\section{Acknowledgment}

The author thanks the Laboratory of Information Technologies 
of JINR (Dubna, Russia) for providing access to its computational
 resources.

\section{Appendix}

\subsection{Kinematic determinants}

The modified Cayley and the Gram determinants
occurring in many formulae of the paper are defined as
\begin{equation}
\Delta_n \equiv \Delta_n(\{p_1,m_1\},\ldots \{p_n,m_n\})=  \left|
\begin{array}{cccc}
Y_{11}  & Y_{12}  &\ldots & Y_{1n} \\
Y_{12}  & Y_{22}  &\ldots & Y_{2n} \\
\vdots  & \vdots  &\ddots & \vdots \\
Y_{1n}  & Y_{2n}  &\ldots & Y_{nn}
\end{array}
         \right|,~~~~~~~~~~~~~~~~~~~~~~
\label{deltan}
\end{equation}
\begin{equation}
Y_{ij}=m_i^2+m_j^2-s_{ij},
\end{equation}
\begin{eqnarray}
G_{n-1} \equiv G_{n-1}(p_1,\ldots ,p_n)= 
-2\left|
\begin{array}{cccc}
  \! S_{11}  & S_{12}  &\ldots & S_{1~ n-1}  \\
  \! S_{21}  & S_{22} &\ldots & S_{2~ n-1} \\
  \vdots  & \vdots  &\ddots & \vdots \\
  \! S_{n-1~ 1}  & S_{n-1 ~2} &\ldots & S_{n-1~ n-1}
\end{array}
\right|, ~~~
\end{eqnarray}
\begin{equation}
~S_{ij}=s_{i n}+s_{jn}-s_{ij},
\label{Gn}
\end{equation}
where
$s_{ij}^2=(p_i-p_j)^2$, 
$p_i$ are combinations of external momenta flowing through $i$-th
lines, respectively, and $m_i$ is the mass of the $i$-th line.
We will use throughout the article an indexed notation for $\Delta_n$ and $G_{n-1}$
\begin{eqnarray}
&&\lambda_{i_1 i_2 \ldots i_n } = 
 \Delta_n(\{p_{i_1},m_{i_1}\},\{p_{i_2},m_{i_2} \},
  \ldots ,\{p_{i_n},m_{i_n}\}),                         \nonumber \\
&& \nonumber \\
&&g_{i_1 i_2 \ldots i_n }=G_{n-1}(p_{i_1},p_{i_2},\ldots ,p_{i_n}).
\label{lage}
\end{eqnarray}
Our results  depend on the ratios of 
$\lambda_{i_1 i_2 \ldots i_n}$  and $g_{i_1 i_2 \ldots i_n}$  and,
therefore,  it is  convenient to introduce the notation
\begin{equation}
r_{ij\ldots k}=-\frac{\lambda_{ij\ldots k}}{g_{ij\ldots k}}.
\label{r_definition}
\end{equation}
Coefficients in front of the integrals in reduction formulae
are expressed in terms of  derivatives of $r_{i ... k}$ with respect
to masses. For convenience we use the following
 shorthand notation,
\begin{equation}
\kappa_{j_r j_1...j_{r-1} j_{r+1}...j_{n}}=
\frac{\partial r_{j_1...j_r ...j_n}}
{\partial m^2_{j_r}}.
\end{equation}

The imaginary part of  $r$ is rather simple. Using
\begin{equation}
\sum_{j=1}^n \partial_j \lambda_{i_1{\ldots} i_n}
= -g_{i_1{\ldots} i_n}= -G_{n-1}(p_{i_1},p_{i_2},\ldots ,p_{i_n}),
\label{Gnm1}
\end{equation}
one shows that to all orders in $\eta$
\begin{equation}
\lambda_{i_1 i_2 \ldots  i_n}(\{m_r^2-i\eta \})=
\lambda_{i_1 i_2 \ldots  i_n}(\{m_r^2\})+i g_{i_1 i_2 \ldots i_n} 
~ \eta,
\end{equation}
and, therefore, the causal $\eta$ prescription for $r$ is
(with the same $\eta$ for all masses)
\begin{equation}
\left. r_{ij\ldots k}\right|_{m_j^2-i\eta}=
\left. r_{ij\ldots k}\right|_{m_j^2}-i\eta.
\label{repsilon}
\end{equation}


\subsection{Hypergeometric functions for the integrals
$I_2^{(d)}$\!\!, $I_3^{(d)}$\!\!, $I_4^{(d)}$  }

In this subsection, we provide collection of some formulae
related to different hypergeometric functions which were
encountered in the derivation of some results of the paper.

\subsubsection{Series representation}
Series representation of the Appell function $F_1$ 
\cite{Appell:1926:FHPH}
\begin{equation}
  F_1(\alpha,\beta,\beta',\gamma;x,y)=\sum_{m,n=0}^{\infty}
  \frac{(\alpha)_{m+n}(\beta)_{m}(\beta')_{n}}
  {(\gamma)_{m+n}} \frac{x^m}{m!}\frac{y^n}{n!}.
\label{F1series}  
\end{equation}
The Appell function $F_3$ is defined by \cite{Appell:1926:FHPH}
\begin{equation}
    F_3(\alpha, \alpha', \beta,\beta',\gamma;x,y)=
   \sum_{m,n=0}^{\infty}\frac{(\alpha)_{m}(\alpha')_{n}
   (\beta)_{m}(\beta')_{n}}{(\gamma)_{m+n}~ }~
   \frac{x^m y^n}{m! ~n!}.
\end{equation}
The Lauricella-Saran function $F_S$ was introduced in
\cite{Lauricella}, \cite{Saran54}
and it is defined by a triple hypergeometric series
\begin{eqnarray}
&&F_S(\alpha_1,\alpha_2,\alpha_2,
       \beta_1, \beta_2, \beta_3;
      \gamma_1,\gamma_1,\gamma_1; x,y,z) \nonumber \\
&&~~=\sum_{r,m,n=0}^{\infty}
 \frac{(\alpha_1)_r (\alpha_2)_{m+n} (\beta_1)_{r}
 (\beta_2)_m (\beta_3)_n }{(\gamma_1)_{r+m+n}~ }~
 \frac{x^r y^m z^n}{r!~ m!~ n!}.
 \label{FSseries}
\end{eqnarray}
A relation between the hypergeometric Lauricella-Saran functions 
$F_S$ and yet another Lauricella-Saran function $F_N$
\cite{Saran55}
\begin{eqnarray}
&&F_S(\alpha_1,\alpha_2,\alpha_2,
       \beta_1, \beta_2, \beta_3;
      \gamma_1,\gamma_1,\gamma_1; x,y,z) = \nonumber \\
&&~~\frac{z^{\beta_2}}{y^{\beta_2}}
F_N\left(\beta_2,\alpha_1,\alpha_2,\beta_2+\beta_3,\beta_1,
\beta_2+\beta_3; \beta_2+\beta_3,\gamma_1,\gamma_1;1-\frac{z}{y},x,z\right),
\label{FS_via_FN}
\end{eqnarray}
where
\begin{eqnarray}
&&F_N(\alpha_1, \alpha_2, \alpha_3, \beta_1,\beta_2,\beta_1;
\gamma_1, \gamma_2,\gamma_2;x,y,z)=
\nonumber \\
&&~~~~~~~~~~~~~~~\sum_{m,n,p=0}^{\infty}
\frac{(\alpha_1)_m (\alpha_2)_n (\alpha_3)_p (\beta_1)_{m+p}(\beta_2)_n}
{(\gamma_1)_m (\gamma_2)_{n+p}} 
\frac{x^m}{m!} \frac{y^n}{n!} \frac{z^p}{p!}.
\end{eqnarray}
More relations between $F_S$ and $F_N$ functions one can find 
in ref. \cite{Saran55}.

The generalized Kamp\'e de F\'eriet hypergeometric function 
$S_1$ in equation (\ref{I3viaS1}) is defined by a double series
\begin{eqnarray}
S_1\(\al,\ap,\bt,\ga,\de; x,y\)
=
\sum_{m,n=0}^{\infty} \frac{(\al)_{m+n}(\ap)_{m+n}(\bt)_m}
{(\ga)_{m+n}(\de)_m}\, \frac{x^m}{m!} \, \frac{y^n}{n!}. 
\end{eqnarray}
The domain of convergence of this series  $|x|+|y|<1$.
The analytic continuation formula for the function $S_1$,
which was used in the derivation of Eq.~(\ref{I3viaS1}) reads 
\cite{Anastasiou:1999ui}
\begin{eqnarray}
\label{eq:s1_anal_cont}
&&S_1\!\(\al,\ap,\bt,\ga,\de;x,y\) =~~~
\nonumber \\
&&~~~
\frac{\Ga{\ap-\al}\Ga{\ga}}{\Ga{\ga-\al}\Ga{\ap}}
(-y)^{-\al} F_2\!\(\al,\bt,\al+1-\ga,\de,\al+1-\ap;-\frac{x}{y},\frac{1}{y}\)
\nonumber \\
&&\nonumber \\
&& +
\frac{\Ga{\al-\ap}\Ga{\ga}}{\Ga{\ga-\ap}\Ga{\al}}
(-y)^{-\ap} F_2\!\(\ap,\bt,\ap+1-\ga,\de,\ap+1-\al;-\frac{x}{y},\frac{1}{y}\). ~~~
\label{S1_analyt_cont}
\end{eqnarray}

The function $F_3$  can be reduced to the function $F_1$ by 
means of \cite{Erdelyi:1953:HTF1}
\begin{equation}
  F_3(\alpha, \alpha', \beta,\beta',\alpha+\alpha';x,y)=
  (1-y)^{-\beta'} F_1 \left(\alpha,\beta,\beta',\alpha+\alpha';
  x,\frac{y}{y-1}\right),
\end{equation}
Similar reduction formula takes place for the Appell function $F_1$ 
\begin{equation}
F_1\left(a,b,b',b+b'; w,z\right)=(1-z)^{-a} \Fh21\Fwz{a,b}{b+b'}.
\end{equation}

\subsubsection{Integral representations}

Euler's integral representation of the  
hypergeometric Gauss' function $_2F_1$
\begin{equation}
{_2F_1}\(\al,\bt,\ga,x\) = \frac{\Ga{\ga}}{\Ga{\bt}\Ga{\ga-\bt}}
\int_0^1 du  \, u^{\bt-1}
(1-u)^{\ga-\bt-1} (1-u x)^{-\al}.
\label{F21intrep}
\end{equation}
\begin{equation}
\bt>0, \quad \ga-\bt >0.
\end{equation}
Euler's integral representation of the Appell function $F_1$
\begin{eqnarray}
&&
F_1(\al, \bt, \bp, \ga; x,y) 
= 
\nonumber \\
&&~~~~~~~
\frac{\Ga{\ga}}{\Ga{\al}\Ga{\ga-\al}} \int_0^1 du \, 
u^{\al-1}(1-u)^{\ga-\al-1}(1-ux)^{-\bt}(1-uy)^{-\bp}\!\!.~~~~~~~
\end{eqnarray}
Euler's integral representation of the Appell function $F_3$
\begin{eqnarray}
&&F_3\(\al,\ap, \bt, \bp, \ga; x, y \) =
\frac{\Ga{\ga}}{\Ga{\bt} \Ga{\bp}
\Ga{\ga-\bt-\bp}} 
\nonumber \\
&&~~~~~~~~~~~~~~~~~~~~~~
\times \int\!\! \int_{u \geq 0,\, v \geq 0 }^{u+v \leq 1} 
\frac{
du\, dv\, u^{\bt-1} v^{\bp-1} (1-u-v)^{\ga-\bt-\bp-1}}
{
(1-ux)^{\al} (1-vy)^{\ap}},~~~
\end{eqnarray}
\begin{equation}
\label{eq:f3_integral}
{\rm Re}(\bt)>0, \quad {\rm Re}(\bp)>0, \quad {\rm Re}(\ga-\bt-\bp)>0.
\end{equation}
An integral representation of the function $S_1$ \cite{Anastasiou:1999ui}
\begin{eqnarray}
&&
S_1\(\al,\ap,\bt,\ga,\de; x,y\) =
\nonumber \\
&&
\frac{\Ga{\ga}}{\Ga{\al}\Ga{\ga-\al}} \int_0^1 du \, 
u^{\al-1}(1-u)^{\ga-\al-1} F_2(\ap,\bt,1,\de,1; ux,uy)
\hfill \cr
\nonumber \\
&&
= \frac{\Ga{\ga}\Ga{\de}}{\Ga{\al}\Ga{\ga-\al}
\Ga{\bt} \Ga{\de-\bt}}
\nonumber \\
&&
\times \int_0^1 du \int_0^1 dv \, u^{\al-1} v^{\bt-1}
(1-u)^{\ga-\al-1} (1-v)^{\de-\bt-1} (1-uvx-uy)^{-\ap}
\end{eqnarray}

In the derivation of Eq.~(\ref{intrepFS}) the following integral representation of
the Lauricella-Saran function $F_S$ was used \cite{Saran55}
\begin{eqnarray}
&&\frac{\Gamma(\alpha_1)\Gamma(\gamma_1-\alpha_1)}
  {\Gamma(\gamma_1)}F_S(\alpha_1,\alpha_2,\alpha_2,
  \beta_1,\beta_2,\beta_3; \gamma_1,\gamma_1,\gamma_1;x,y,z)
  = \nonumber\\
&& \nonumber \\
&&~~\int_0^1 \frac{t^{\gamma_1-\alpha_1-1} (1-t)^{\alpha_1-1}}
  {(1-x+tx)^{\beta_1}}F_1(\alpha_2,\beta_2,\beta_3,
  \gamma_1-\alpha_1;ty,tz) dt.
\label{FSintF1}
\end{eqnarray}

\subsection{Differential relations for the Lauricella-Saran
function $F_S$}
In order to obtain system of the differential equations for the 
boundary function $C_4(x,y,z)$ in equation
(\ref{I4solution_dr}), we used the following
differential relations for the Lauricella-Saran function
$F_S$:
\begin{eqnarray}
&&
2 (1-x) x (x+y-x y) (x+z-x z)
\frac{\partial}{\partial x}F_s^{(d)}(x,y,z)=
\nonumber \\
&&x^2 (d-2)
-x^2 \Fh21\Fx{1,\frac{d-3}{2}}{\frac{d}{2}}
+(1-x) x z (d-3)\Fh21\Fz{\frac12, 1}{\frac{d}{2}}
\nonumber \\
&&
+y (d-4) (x z-x-z) (x-1)F_1\left(1,1,\frac12,\frac{d}{2};y,z\right)
\nonumber \\
&&
+(d x y-d x-d y-3 x y+3 x+4 y) (x z-x-z) (x-1)F_s^{(d)}(x,y,z)
,
\nonumber 
\end{eqnarray}
\vspace{0.3cm}
\begin{eqnarray}
&&
2 (1-y) (y-z)(x+y-x y)
\frac{\partial}{\partial y}F_s^{(d)}(x,y,z)=
x (1-y)\Fh21\Fx{1, \frac{d-3}{2}}{\frac{d}{2}}
\nonumber \\
&&
-z (d-3) \Fh21\Fz{\frac12, 1}{\frac{d}{2}}
-(y-1) (2xy-x z-x-2y+z) F_s^{(d)}(x,y,z)
\nonumber \\
&&-(y-z)(d-4)F_1\left(1,1,\frac12,\frac{d}{2};y,z\right)
+y (d-2),
\nonumber 
\end{eqnarray}
\begin{eqnarray}
&&
2 (1-z) (y-z) (x+z-x z)
\frac{\partial}{\partial z}F_s^{(d)}(x,y,z)=
x (z-1)\Fh21\Fx{1,\frac{d-3}{2}}{\frac{d}{2}}
\nonumber 
\\
&&
+z (d-3)\Fh21\Fz{1,\frac12}{\frac{d}{2}}
+(z-1) (x z-x-z)F_s^{(d)}(x,y,z)-z (d-2),
\end{eqnarray}
where
\begin{eqnarray}
&&F_s^{(d)}(x,y,z)=
F_S\left(\frac{d-3}{2},1,1,1,1,\frac12;\frac{d}{2},\frac{d}{2},\frac{d}{2};
x,y,z\right).
\end{eqnarray}
These differential relations were derived by using series representation
(\ref{FSseries}).

Solving dimensional recurrence relation for the integral $I_4^{(d)}$
we used the following recurrence relation 
for the hypergeometric function $F_S$
\begin{equation}
(d-3) xF_s^{(d+2)}(x,y,z)=d F_s^{(d)}(x,y,z)
-d F_1\left(1,1,\frac12,\frac{d}{2};y,z\right).
\end{equation}

Additionally, we provide here differential relations for the Appell function
$F_1$, which were used to find system of differential 
equations for the function $C_4(x,y,z)$ 
\begin{eqnarray}
&&
\frac{\partial}{\partial x}
F_1\left(\frac{d-3}{2},1,\frac12,\frac{d-1}{2}; x,y\right)
=- \frac{ (dx-dy-3x+4y)}{2x(x-y)} 
F_1\left(\frac{d-3}{2},1,\frac12,\frac{d-1}{2}; x,y\right)
\nonumber \\
&&~~~~~~~~~~~ -\frac{y(d-4)}{2x(x-y)}
\Fh21\Fy{\frac12, \frac{d-3}{2}}{\frac{d-1}{2}}
+\frac{(d-3)\sqrt{1-y}}{2(x-y)(1-x)},
\nonumber \\
&& \nonumber \\
&&
\frac{\partial}{\partial y}
F_1\left(\frac{d-3}{2},1,\frac12,\frac{d-1}{2}; x,y\right)
= \frac{1}{2(x-y)} 
F_1\left(\frac{d-3}{2},1,\frac12,\frac{d-1}{2}; x,y\right)
\nonumber \\
&&~~~~~~~~~~
+\frac{(d-4)}{2(x-y)}\Fh21\Fy{\frac12,\frac{d-3}{2}}{\frac{d-1}{2}} 
-\frac{d-3}{2(x-y)\sqrt{1-y}  }
\end{eqnarray}
In order to obtain these relations we used series representation (\ref{F1series}).


\end{document}